\documentclass[12pt,preprint]{aastex}

\usepackage{lscape}
\usepackage{subfigmat}
\usepackage{rotating}

\newcommand{\chisq}{$\chi^{2}$}
\newcommand{\eqnname}{Eqn.~}
\newcommand{\figname}{Figure~}

\newcommand{\tabname}{Table~}
\newcommand{\app}{Appendix~}

\newcommand{\breakfourthousand}{4000~\AA~break}
\newcommand{\fourthousandbreak}{4000~\AA~break}

\newcommand{\bindex}{\dn}
\newcommand{\dn}{$\mathrm{D}_{n}(4000)$}  
\newcommand{\gband}{G~band}

\newcommand{\hdeltaa}{$\mathrm{H}{\delta_{A}}$}
\newcommand{\hgammaa}{$\mathrm{H}{\gamma_{A}}$}
\newcommand{\hdeltaf}{$\mathrm{H}{\delta_{F}}$}
\newcommand{\hgammaf}{$\mathrm{H}{\gamma_{F}}$}
\newcommand{\hbeta}{H$\beta$}
\newcommand{\hdelta}{H$\delta$}
\newcommand{\halpha}{H$\alpha$}
\newcommand{\hgamma}{H$\gamma$}

\newcommand{\nad}{Na\,D}
\newcommand{\woindices}{\hdeltaa, \hgammaa, \hdeltaf, \hgammaf}

\newcommand{\metallicity}{$Z$}
\newcommand{\nmetallicity}{6}
\newcommand{\nagegiven}{221}
\newcommand{\nage}{182}
\newcommand{\nspec}{1092}
\newcommand{\nlickfirst}{21}
\newcommand{\nlicksecond}{4}
\newcommand{\nlickthird}{1}
\newcommand{\ssp}{SSP}
\newcommand{\lick}{Lick/IDS}
\newcommand{\lickconfres}{9~\AA}
\newcommand{\lickconfresunit}{FWHM}
\newcommand{\lickconfsampling}{1.25~\AA \ per pixel}
\newcommand{\lickconfwrange}{3800--6400~\AA}
\newcommand{\lickwrange}{4000--6400~\AA}
\newcommand{\sdssspecres}{1800}
\newcommand{\sdssspecreskms}{166~km~s$^{-1}$}
\newcommand{\sdssspecreswc}{3450~\AA}
\newcommand{\sdssspecresang}{1.9~\AA}
\newcommand{\sdssspecobswave}{3800--9200~\AA}
\newcommand{\sdssgalmedianredshift}{0.1}
\newcommand{\bcmodelres}{3~\AA}
\newcommand{\bcmodelresunit}{FWHM}
\newcommand{\sdssconfres}{\bcmodelres}
\newcommand{\sdssconfresunit}{FWHM}
\newcommand{\sdssconfsampling}{69~km~s$^{-1}$ per pixel}
\newcommand{\sdssconfwrange}{3450--8350~\AA}

\newcommand{\anrow}{m}
\newcommand{\ancol}{n}
\newcommand{\cutoutncol}{n^{(i)}}

\newcommand{\cur}{CUR Matrix Decomposition}
\newcommand{\curonly}{CUR}
\newcommand{\costheta}{$\costhetamath$}
\newcommand{\subspaceangle}{GNS}
\newcommand{\costhetamath}{\cos\theta_\mathrm{\subspaceangle}}
\newcommand{\cossqtheta}{$\cos^{2}\theta_\mathrm{\subspaceangle}$}
\newcommand{\subfirstpcangle}{PC1}
\newcommand{\cossqthetafirstpc}{$\cos^{2}\theta_\mathrm{\subfirstpcangle}$}
\newcommand{\costhetafirstpc}{$\cos\theta_\mathrm{\subfirstpcangle}$}
\newcommand{\datamatrix}{$A$}
\newcommand{\datavalue}{f_{\lambda} \cdot \Delta{\lambda}}
\newcommand{\datavalueij}{f^{i}_{\lambda_{j}} \cdot \Delta{\lambda_{j}}}

\newcommand{\extendedlick}{Extended~\lick~indices}
\newcommand{\idspace}{object~space}
\newcommand{\leftsingularvectorone}{$u^{1}$}
\newcommand{\leftsingularvectortwo}{$u^{2}$}
\newcommand{\varspace}{variable~space}
\newcommand{\name}{Region}
\newcommand{\nextendedlick}{26}
\newcommand{\neigen}{$k$}

\newcommand{\parecoeff}{parameter-eigencoefficient}
\newcommand{\parvector}{$p$}
\newcommand{\pearsoncoeff}{$r$}
\newcommand{\pearsoncoeffsq}{$r^2$}
\newcommand{\pca}{PCA}
\newcommand{\pixfirst}{1}
\newcommand{\pixlast}{N}
\newcommand{\pj}{$p_{\lambda}$}
\newcommand{\pjsum}{$p^{S}_{\lambda}$}

\newcommand{\pjsummath}{p^{S}_{\lambda}}
\newcommand{\pjsumregion}{p^{S}}
\newcommand{\pjsumthreshold}{$T$}
\newcommand{\pjname}{\textit{LeverageScore}}
\newcommand{\pjsumname}{\textit{LeverageScoreSum}}

\newcommand{\roi}{ROI}

\newcommand{\roiws}{\lambda_{S}}
\newcommand{\roiwe}{\lambda_{E}}
\newcommand{\roiwave}{[$\roiws, \roiwe$]}

\newcommand{\realspace}{\mathbb{R}}
\newcommand{\region}{region}
\newcommand{\score}{Leverage Score}
\newcommand{\svd}{SVD}

\newcommand{\wavec}{$\lambda_{I}$}

\newcommand{\chosenpjsumthreshold}{0.7}
\newcommand{\chosenpjsumthresholdsd}{0.1}
\newcommand{\datareduction}{$\approx100$}
\newcommand{\firstpar}{stellar age}
\newcommand{\secondpar}{stellar metallicity}
\newcommand{\firstespecshows}{Balmer~absorption}
\newcommand{\firstespec}{stellar age}
\newcommand{\secondespecshows}{\fourthousandbreak}
\newcommand{\secondespec}{age-metallicity degeneracy}
\newcommand{\thirdespec}{anti-correlation between the strengths of the Balmer and the Ca~K \& H absorptions}
\newcommand{\firstnsdssconf}{15}
\newcommand{\firstnlickconf}{15}
\newcommand{\firstncompare}{12}
\newcommand{\linereconacc}{$\approx{20}\%$}
\newcommand{\manypar}{$>10$}
\newcommand{\widthidconvergencetest}{NaD}

\newcommand{\lickmostinformative}{\dn, \hbeta \ and \hdeltaa}
\newcommand{\lickconfkpick}{10}

\newcommand{\lickconffirstimportantregion}{\breakfourthousand}
\newcommand{\lickconfsecondimportantregion}{\hbeta}
\newcommand{\lickconfthirdimportantregion}{\hdelta}
\newcommand{\lickconffourthimportantregion}{Fe4531}

\newcommand{\lickconfunimportantregion}{5500~\AA}
\newcommand{\lickconfminregions}{three}
\newcommand{\sdssconfkpick}{25}
\newcommand{\sdssconfwidthnad}{14.9~\AA}
\newcommand{\sdssconffirstsecondpjsumratio}{6}
\newcommand{\sdssconffirstimportantregion}{\fourthousandbreak \ and the \hdelta \ line}
\newcommand{\sdssconffirstimportantregionname}{\name{01}}
\newcommand{\sdssconfsecondimportantregion}{Fe-like indices}
\newcommand{\sdssconfthirdimportantregion}{\hbeta \ line}
\newcommand{\sdssconffourthimportantregion}{\gband \ and the \hgamma \ line}
\newcommand{\sdssconfpjsumcomparableregions}{2nd, 3rd,  4th}
\newcommand{\sdssconfpjsumcomparableamp}{0.05}
\newcommand{\sdssconfhalpharegionorder}{7th}
\newcommand{\sdssconfhalphaimportantregion}{\halpha}

\newcommand{\sdssconfminregions}{first \region}
\newcommand{\pjsumcutoff}{0.03}

\begin{document}

 \title{Objective Identification of  Informative Wavelength Regions in
   Galaxy Spectra}

\author{Ching-Wa~Yip\altaffilmark{1},
  Michael~W.~Mahoney\altaffilmark{2},                  
  Alex~S.~Szalay\altaffilmark{1,3}, 
  Istv\'an~Csabai\altaffilmark{4},
  Tam\'as~Budav\'ari\altaffilmark{1},
  Rosemary~F.~G.~Wyse\altaffilmark{1}, 
  Laszlo~Dobos\altaffilmark{4}}

\altaffiltext{1}{Department  of  Physics   and  Astronomy,  The  Johns
  Hopkins  University, 3701  San  Martin Drive,  Baltimore, MD  21218,
  USA.}    
\altaffiltext{2}{Department    of   Mathematics,   Stanford
  University,  Stanford, CA 94305,  USA.}  
\altaffiltext{3}{Department
  of Computer  Science, The Johns Hopkins University,  3400 N. Charles
  Street, Baltimore, MD 21218, USA.}  
\altaffiltext{4}{Department   of   Physics   of   Complex   Systems,
  E\"{o}tv\"{o}s Lor\'and University, H-1117 Budapest, Hungary.}

\email{cwyip@pha.jhu.edu;mmahoney@cs.stanford.edu;szalay@jhu.edu}

\begin{abstract}
Understanding the diversity  in spectra is the key  to determining the
physical parameters  of galaxies. The optical spectra  of galaxies are
highly  convoluted  with continuum  and  lines  which are  potentially
sensitive  to different physical  parameters. Defining  the wavelength
\region{s} of  interest is therefore  an important question.   In this
work, we identify informative  wavelength \region{s} in a single-burst
stellar  populations model by  using the  \cur.  Simulating  the \lick
\   spectrograph   configuration,   we   recover   the   widely   used
\lickmostinformative \  to be  most informative.  Simulating  the SDSS
spectrograph  configuration with  a  wavelength range  \sdssconfwrange
\ and  a model-limited spectral  resolution of \sdssconfres,  the most
informative       \region{s}       are:       first       \region--the
\sdssconffirstimportantregion;           second           \region--the
\sdssconfsecondimportantregion;           third           \region--the
\sdssconfthirdimportantregion;           fourth           \region--the
\sdssconffourthimportantregion.  A Principal Component Analysis on the
first \region \ shows that the first eigenspectrum tells primarily the
\firstespec, the second eigenspectrum  is related to the \secondespec,
and the third eigenspectrum  shows an \thirdespec.  The \region{s} can
be used  to determine  the stellar age  and metallicity  in early-type
galaxies  which   have  solar  abundance   ratios,  no  dust,   and  a
single-burst  star formation  history.  The  \region  \ identification
method can be applied to any set of spectra of the user's interest, so
that  we  eliminate the  need  for  a  common, fixed-resolution  index
system.   We  discuss  future  directions  in  extending  the  current
analysis to late-type galaxies.
\end{abstract}


\section{Introduction}\label{section:introduction}

It has  long been known that  the diversity in spectra  of galaxies is
driven by  the variance in their underlying  physical properties, such
as stellar age  and metallicity.  The line indices,  or the relatively
narrow  regions  of  a   spectrum,  are  isolated  for  the  parameter
sensitivities.        \citet{1973ApJ...179..731F}      found      that
metallicity-sensitive  indices  from  a 10-filter  photometric  system
correlate  with  the  luminosity  of  ellipticals,  giving  incredible
insights to the  role of gravitational potential well  to the chemical
evolution   in  those  galaxies.    The  concept   of  ``informative''
wavelength regions  had since been extended  to spectroscopy.  Perhaps
the    most    well-known    example   is    the    \breakfourthousand
\ \citep{1983ApJ...273..105B} and its correlation with the stellar age
and  metallicity in early-type  galaxies.  \citet{1994ApJS...94..687W}
later assembled \nlickfirst \ absorption indices and measured them for
several hundred stars in the Galaxy, and the same set were measured by
\citet{1998ApJS..116....1T} for several  hundred globular clusters and
galaxies.      They    are    called     the    \lick     \    indices
\citep[][]{1984ApJ...287..586B,1985ApJS...57..711F}.    These  indices
are continuum subtracted,  where the continua are defined  on the left
and right flanks in the vicinity of the actual index wavelength range.
Within an index  wavelength range, one or more  absorption lines, or a
feature,  are  present.   Therefore,  the  index  strength  can  be  a
sensitive   measure   for  specific   galaxy   parameters.   The   \dn
\citep{1999ApJ...527...54B}\footnote{Renamed                         by
  \citet{2003MNRAS.341...33K}  to be ``\dn'',  with the  subscript $n$
  presumably signifying the left  and right regions of influence being
  narrower       than      the       previous       definition      by
  \citet{1983ApJ...273..105B}.       The     name      appeared     in
  \citet{1999ApJ...527...54B}    is   ``D(4000)''.}     and   \hdeltaa
\ \citep{1997ApJS..111..377W}  indices were subsequently  defined, and
were used to constrain stellar  age and the starburst mass fraction in
galaxies \citep{2003MNRAS.341...33K}.

The concept of informative wavelength \region{s} has also been used in
conjunction with  a popular data compression  technique, the Principal
Component Analysis (\pca), to  derive physical parameters of galaxies.
Each spectrum in a galaxy sample is represented by the weighted sum of
the    same    orthogonal    basis,   hereafter    the    eigenspectra
\citep{1995AJ....110.1071C},   and   the   weights  are   called   the
eigencoefficients. The  lowest orders of  the eigenspectra encapsulate
most     of    the    sample     variance    in     nearby    galaxies
\citep[][]{2004AJ....128..585Y}, forming  a {\it subspace}  which lies
in    a   higher-dimensional    wavelength    space.    Using    \pca,
\citet{2007MNRAS.381..543W}  have measured, for  a galaxy  sample from
the SDSS \citep{2000AJ....120.1579Y}, the eigencoefficients within the
restframe 3750--4150~\AA  \ out of  the full optical  wavelength range
available,   3800--9200~\AA   \    in   the   observed   frame.    The
eigencoefficients were  used to correlate with  physical properties of
the  galaxies.    Later,  \citet{2012MNRAS.421..314C}  have  estimated
model-based stellar  populations and dust  parameters for a  sample of
SDSS  galaxies.    They  used  eigencoefficients   calculated  in  the
3700--5500~\AA  \ range,  in conjunction  with the  Bayesian parameter
estimation method.

Despite decades of effort and  progress, the \lick \ indices remain to
be subjective. The root cause lies in the difficulties in defining the
true continua in the vicinity of a feature.  Two main hurdles are: (1)
There  is  no ``true''  continuum  in  late-type  stars at  the  \lick
\    spectral   resolution   and    the   continua    are   ``pseudo''
\citep{1994ApJS...94..687W}.  (2)  The location  and the width  of the
pseudocontinua are  subjective. The location  are chosen to  flank the
feature, and  the width to  be large enough  in order to  minimize the
effect  of typical  stellar  velocity dispersion  of  galaxies on  the
measured                        index                        strengths
\citep{1994ApJS...94..687W,1998ApJS..116....1T}.   For a  galaxy  of a
given  velocity   dispersion,  these  characteristics   determine  the
absorption lines present within the pseudocontinua, in turn impact the
parameter  sensitivity  of the  corresponding  feature.  For  example,
\citet{1995ApJ...446L..31J}   found    in   a   single-burst   stellar
populations model that the narrow Balmer index, \hgamma$_\mathrm{HR}$,
is  significantly  more  sensitive  to  stellar  age  than  its  broad
counterpart, \hgammaa, suggesting that  the broader Balmer indices may
be   contaminated  by   metal  lines.    They  are   the   Fe~I  lines
\citep{2004MNRAS.351L..19T},  present  in  the pseudocontinua  of  the
broad, higher-order Balmer indices \citep{1997ApJS..111..377W}.

   The \pca \ approach \citep{2007MNRAS.381..543W,2012MNRAS.421..314C}
   abandoned the  use of pseudocontinua and  bypassed difficulties (1)
   \& (2), but the  feature wavelength ranges were chosen subjectively
   to include the \fourthousandbreak,  unlike the \lick \ index system
   in which they were measured.   Like the case in pseudocontinua, the
   identity  and the  width  of  a feature  can  impact its  parameter
   sensitivity,  because  they  determine  what absorption  lines  are
   present  within.   The  reason  for   using  \pca  \  is  that  the
   eigenspectra  are   orthogonal,  so  that   each  eigenspectrum  is
   potentially sensitive to  individual physical parameters.  Clearly,
   the     next-generation    approach     is    to     combine    the
   orthogonal-decomposition  strength of  \pca \  with  objectivity in
   defining the feature wavelength \region{s}.

The  main  goals  of  this  paper  are  to  identify  the  informative
wavelength  \region{s}  in  an  objective  manner,  and  to  associate
physical  significance with  them.  The  analyses are  performed  on a
solar-abundance model of Simple Stellar Populations that is defined by
two  parameters,  stellar  age  and  metallicity.   We  use  the  \cur
\   \citep{2008Drineasetal,2009MahoneyDrineas},  a   powerful  machine
learning  technique  which has  the  ability  to select  statistically
informative columns  from a data  matrix.  The key  characteristics of
our  approach are: (a)  The continua  are not  explicitly used  in the
\region  \ identification,  bypassing the  above-mentioned (1)  \& (2)
which plague  the \lick  \ indices.  (b)  Different from  the previous
\pca  \  approach \citep{2007MNRAS.381..543W,2012MNRAS.421..314C},  in
this work the \region{s} are  determined objectively, in terms of both
the identity and the width.

We  cast  an  eye  towards  estimating  physical  parameters  of  real
galaxies. This  work results in two related  applications.  First, the
identified \region{s}  and the  \parecoeff \ relation  can be  used to
determine the stellar age and metallicity of those early-type galaxies
which have  solar abundance ratios,  no dust, and a  single-burst star
formation history.  Second, our \region \ identification method can be
applied to  any set  of model spectra.   Users can generate  their own
\parecoeff  \  relation from  the  model,  and  use that  to  estimate
parameters  of galaxies  from  the observed  spectra.  Therefore,  the
intricate processes \citep{1997ApJS..111..377W} involved in conforming
to a common index system,  which attempt to account for the difference
in spectral  resolution between the  observed spectra and  the system,
can be eliminated.

In   \S\ref{section:data},   we  present   the   model  spectra.    In
\S\ref{section:analysis}, we present the \cur \ and describe how it is
used to identify informative  wavelength regions in the model spectra.
In \S\ref{section:results}, we  present the identified \region{s}, the
comparison  to   the  existing   line  indices,  and   their  physical
significance.   In \S\ref{section:summary}, we  conclude our  work and
discuss next steps.  Vacuum wavelengths are used throughout the paper.

\section{Model \& Preprocessing}\label{section:data}

We  wish to  derive a  set  of informative  \region{s} applicable  for
estimating the stellar  age and metallicity in galaxies  which have no
dust  and follow  a  single-burst star  formation  history.  Hence  we
perform the  \cur \  on a model  with known physical  parameters.  The
\cite{2003MNRAS.344.1000B} stellar  populations model is  used.  For a
chosen stellar initial mass function (IMF) and stellar isochrones, the
model  provides  spectra   of  Simple  Stellar  Populations  (\ssp{s})
spanning \nmetallicity  \ stellar metallicities  by mass (\metallicity
\ = 0.0001, 0.0004, 0.004, 0.008, 0.02, 0.05) and \nagegiven \ stellar
ages (0  -- 20~Gyr,  distributed roughly in  equally-sized logarithmic
bins).  We  adopt the \citet{2003PASP..115..763C}  IMF and Padova~1994
isochrones \citep[][and  references therein]{1996A&AS..117..113G}, and
these choices  are not expected  to impact our  results qualitatively.
We  do not  add extinction  to our  model.  And,  there is  no stellar
velocity dispersion variance  in our model, in contrast  to the galaxy
spectra     used      the     \lick     \      indices     measurement
\citep{1998ApJS..116....1T}.   We choose  not to  study the  effect of
noise and other artifacts  on the identified \region{s}, because other
approaches \citep[e.g.,  Robust \pca \  by][]{2009MNRAS.394.1496B} can
be   used  to   derive  robust   eigenspectra   and  eigencoefficients
\citep{2012MNRAS.420.1217D}  for \region{s}  in  real galaxy  spectra.
The \cite{2003MNRAS.344.1000B} model has solar abundance ratios, which
may  limit the  application  of our  \region{s}  from many  early-type
galaxies \citep[e.g.,][]{1992ApJ...398...69W} having non-solar values.
We decide to identify a new set of \region{s} when a more sophisciated
stellar populations model is available in the future.


A  preprocessing is  carried out  to ensure  that (i)  each  and every
spectrum is  unique\footnote{Some \ssp \  spectra are found to  be the
  same  in  the  \chisq \  sense,  despite  the  fact that  they  have
  different ages. For  a given metallicity, we retain  the spectrum at
  the earlier age when a pair of spectra at two {\it consecutive} ages
  are  the same.  The  real degeneracies  between age  and metallicity
  still remain.}, (ii) all of  the flux values are available (i.e., no
NaN, and no  zero vector), and (iii) the  final ages and metallicities
are selected as  such they follow a rectangular  parameter grid.  This
step  results in  the  same \nmetallicity  \  metallicities but  \nage
\ ages (2.5~Myr -- 20~Gyr), or \nspec \ \ssp \ spectra in total.

 The  \ssp  \ are  rebinned,  in  a  flux-conserving fashion,  to  two
 configurations: the \lick  \ and the SDSS. The  \lick \ configuration
 allows us  to compare  the identified \region{s}  to the  widely used
 indices in the literature.  On the other hand, the SDSS configuration
 will result in \region{s} with  a spectral resolution higher than the
 \lick \  indices and  closer to the  SDSS galaxy spectra.   The \lick
 \  configuration is  a wavelength  range \lickconfwrange,  a spectral
 resolution  of  \lickconfres~\lickconfresunit,   and  a  sampling  of
 \lickconfsampling.  The  model spectra  are therefore scaled  down in
 resolution by convolving with a Gaussian function. The function has a
 FWHM    equal   to    the    \citet{2003MNRAS.344.1000B}   resolution
 (\bcmodelres~\bcmodelresunit      \      within     the      optical)
 quadrature-subtracted from  the targeted resolution.   This choice is
 made    to   mimic   the    \lick   \    spectrograph   configuration
 \citep[\lickwrange  \ in air  wavelengths,][]{1984ApJ...287..586B} as
 well as  to include both the left  and the right flanks  of \dn. They
 are  referred as  the \dn{L}  and \dn{R}  for convenience.   The SDSS
 configuration  is  a  wavelength  range \sdssconfwrange,  a  spectral
 resolution  of  \sdssconfres~\sdssconfresunit,   and  a  sampling  of
 \sdssconfsampling.  The  choice of \sdssconfres  \ is limited  by the
 instrumental  resolution  of  the \citet{2003MNRAS.344.1000B}  model.
 The  actual   spectral  resolution   in  the  SDSS   is  \sdssspecres
 \ (corresponds to \sdssspecreskms), higher  than that of the model at
 the  shortest wavelengths, and  is \sdssspecresang  \ at  $\lambda =$
 \sdssspecreswc.  The  wavelength range is tuned to  coincide with the
 restframe wavelength  range of spectra in most  nearby SDSS galaxies,
 for  which  the   median  redshift  is  \sdssgalmedianredshift.   The
 observed-frame wavelengths  of the SDSS spectrograph  are ranged from
 \sdssspecobswave.


The    spectra     are    converted    from     flux    density    (in
erg~s$^{-1}$~\AA$^{-1}$)    to    flux    (in   erg~s$^{-1}$)    using
$f_{\lambda}^{\prime} =  \datavalue$.  The data values,  which will be
cast into a matrix later, are  in the flux unit.  The mean spectrum of
the  model is subtracted  from each  \ssp \  spectrum. The  data cloud
formed by the spectra are hence centered in the wavelength space.  The
spectra are  not continuum  subtracted, as such  no assumption  on the
true continuum is made.

\subsection{\extendedlick}

We  assemble the  \nlickfirst \  \lick \  indices  from \citet[][their
  whole  \tabname2]{1998ApJS..116....1T}, the  \nlicksecond  \ indices
from       \citet[][\woindices       \       or      their       whole
  \tabname1]{1997ApJS..111..377W}  and the  \nlickthird  \ index  from
\citet[][\bindex  \ in  \tabname1]{1999ApJ...527...54B}, and  name the
\nextendedlick   \   collectively   as   ``the   \extendedlick''   for
convenience.  The details are listed \tabname\ref{tab:lick}.

\section{Analysis}\label{section:analysis}

\subsection{\cur} 

The  \cur  \  \citep{2008Drineasetal,2009MahoneyDrineas}  is  a  novel
method that  has been used  for large-scale data analysis  in multiple
disciplines.  The main idea is  to approximate a potentially huge data
matrix with a lower rank matrix, where the latter is made with a small
number  of \emph{actual}  rows  and/or columns  of  the original  data
matrix.   The selected  rows and  columns are  therefore statistically
informative.  The data  in this work are model  spectra, selecting the
rows  and  columns  corresponds  to  selecting the  spectrum  IDs  and
wavelengths, respectively.   We will use  only the column part  of the
decomposition,  selecting  informative  wavelengths  in  the  spectra.
Further    details    of    the    \cur    \    are    presented    in
\app\ref{appendix:cur}.

We  start  by  constructing  a  $\anrow  \times  \ancol$  data  matrix
\datamatrix, where each row is a single \ssp \ spectrum from the model
spectra.  That  is, $A_{ij} = \datavalueij$  is the flux  of the $i$th
spectrum  at the  $j$th wavelength.   The  full set  of model  spectra
(\S\ref{section:data})  is used, so  that the  sample variance  in the
matrix  is driven  by both  stellar age  and metallicity.   A Singular
Value Decomposition (\svd) is performed on \datamatrix
\begin{equation}
A = U \, \Sigma \, V^{T} \ ,
\end{equation}

\noindent
which gives $U$ and $V$, the orthonormal matrices in which the columns
are respectively  the left and  right singular vectors;  and $\Sigma$,
the  diagonal matrix  containing the  singular values.   For  a chosen
\neigen, the \pjname \ at a given wavelength is
\begin{equation}
 p_{\lambda}   =    \frac{\sum_{j   =   1}^{\mathit{k}}    \,   \left(
   v^{j}_{\lambda} \right)^{2}}{k} \ , \label{eqn:pj}
\end{equation}

\noindent
where  $v^{j}_{\lambda}$  is the  $j$th  right  singular vector.   The
collection  of the  $k$ right  singular vectors  forms  an orthonormal
basis  to the fluxes  in the  high-dimensional wavelength  space.  The
larger the $v^{j}_{\lambda}$ value,  the larger the sample variance is
projected from the $j$th basis  vector onto the given wavelength axis.
The \pj \ is hence proportional to the total projected sample variance
from altogether $k$ basis vectors onto that wavelength axis.  In other
words, \pj \ measures the  information contained in a wavelength. This
information  tells   the  sample  variance  driven   by  the  physical
parameters, stellar age and metallicity.

The ordering of  the spectra per row in the  matrix \datamatrix \ does
not impact the ordering of the right singular vectors.  As long as the
same set of spectra is used, and  the same $k$ is chosen, the same \pj
\ will result.  The $k$ values will be chosen  in an objective manner
in this work.

\subsection{Region Identification Procedure}

The two main  steps in the \region \ identification  are to select the
informative  wavelength,  \wavec,  and  to  determine  its  region  of
influence,  \roi.   To  select  \wavec,  we pick  from  the  available
wavelengths the one  with the highest \pj.  To  determine the \roi, we
observe  several  guidelines:  (A)   A  \region  \  is  contiguous  in
wavelength. (B) A \region \  is allowed to be asymmetric about \wavec.
(C) Regions  can overlap with  each other.  (D)  We expect that,  if a
region can  indeed be defined,  the \pj \  would decline to  zero when
approaching the  left and right wavelength bounds,  because the fluxes
Combining the above considerations, a \roi \ is calculated as follows.
Starting at  the wavelength that  is selected (\wavec), we  attempt to
include a pixel to its  immediate left or right.  The accumulated \pj \
of both scenarios are calculated,  called \pjsum.  The pixel gives the
higher \pjsum  \ is added to  the \region.  The  procedure is repeated
until  the last  ($i$th)  pixel  is included,  as  such the  following
convergence criterion is satisfied
\begin{equation}
\left|  \frac{{\pjsummath}_{i} -  {\pjsummath}_{i -  1}}{\lambda_{i} -
  \lambda_{i      -      1}}       \right|      \leq      T      \cdot
\frac{\sum_{\lambda=\lambda_1}^{\lambda_N}
  p_{\lambda}}{\lambda_{\pixlast}  -  \lambda_{\pixfirst}}  = T  \cdot
\frac{1}{\lambda_{\pixlast}           -           \lambda_{\pixfirst}}
\ . \label{eqn:convergence}
\end{equation}

\begin{figure}
\plotone{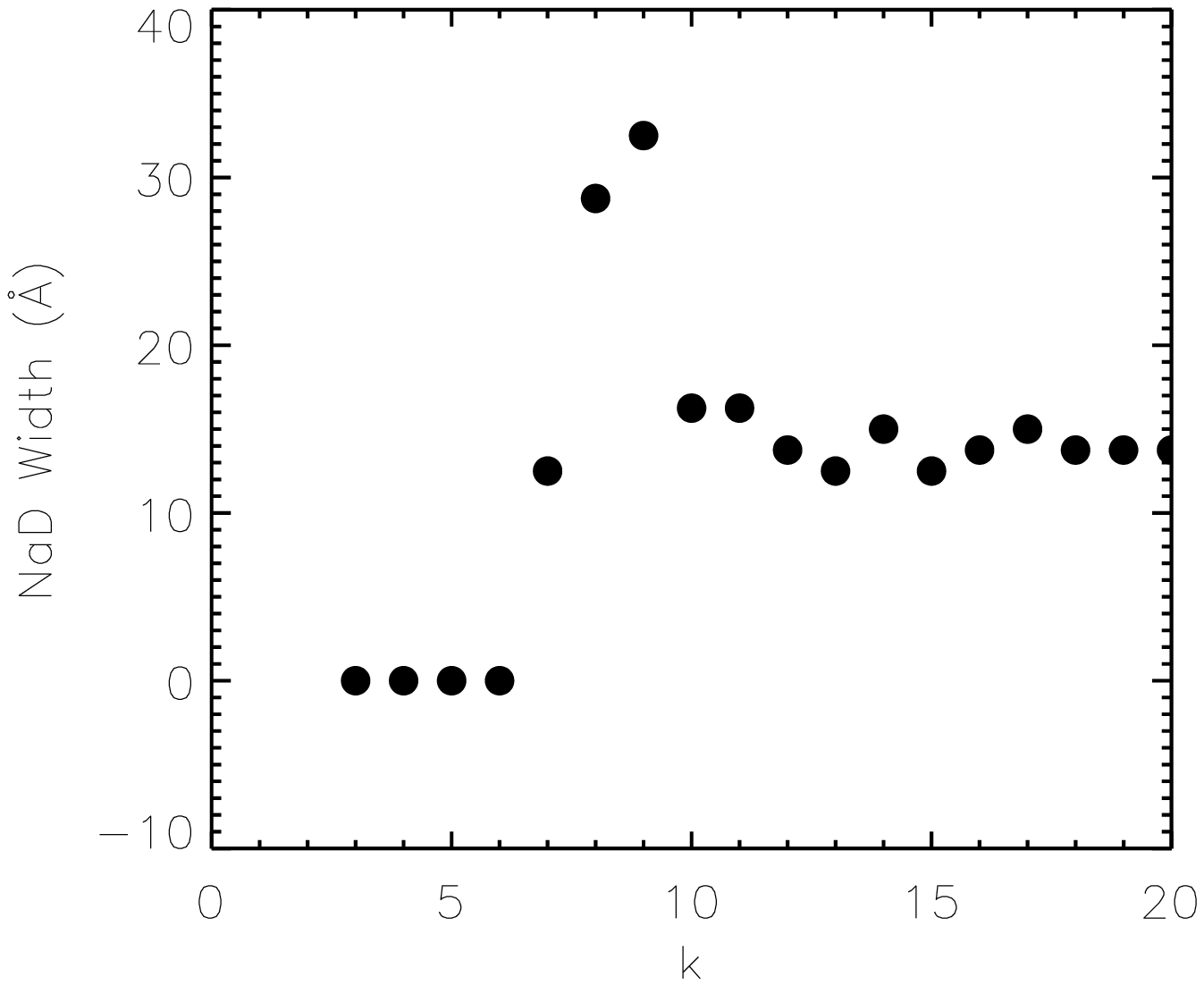}
\caption{The width of the \nad-\region \ as a function of \neigen, for
  \pjsumthreshold   =    \chosenpjsumthreshold   \   in    the   \lick
  \ configuration. The width converges.}
\label{fig:RegionWidth_NaD_vs_k_T0.70_Lick}
\end{figure}

\begin{figure}
\plotone{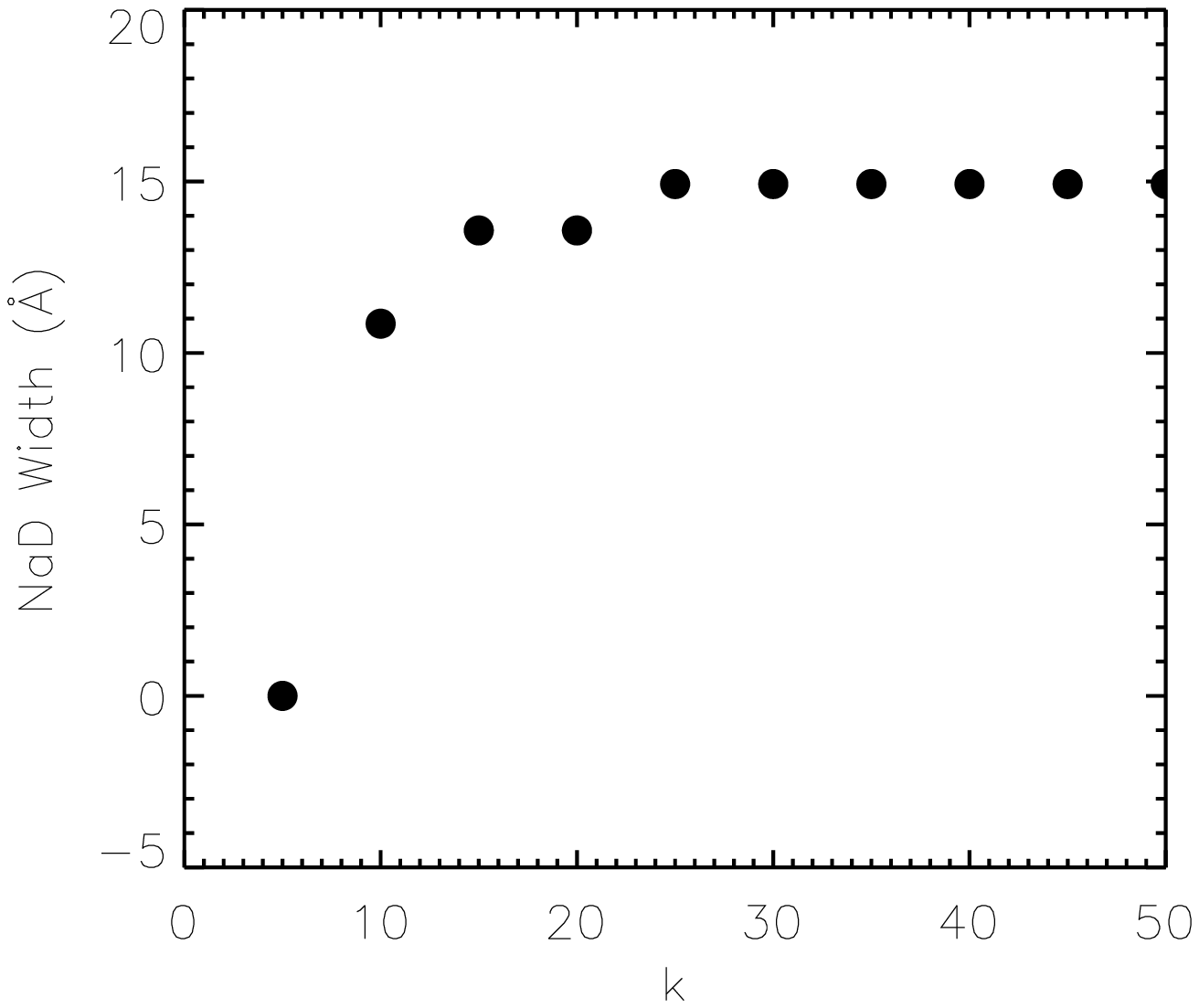}
\caption{The width of the \nad-\region \ as a function of \neigen, for
  \pjsumthreshold    =   \chosenpjsumthreshold    \   in    the   SDSS
  configuration. The width converges.}
\label{fig:RegionWidth_NaD_vs_k_T0.70_SDSS}
\end{figure}

\noindent
That is, the change of the information content of a \region \ per unit
wavelength  is  a  constant.   We  tie the  constant  to  the  average
information  expected in the  case where  every wavelength  is equally
informative.    The   threshold,  \pjsumthreshold,   is   set  to   be
\chosenpjsumthreshold  \  in this  work.   This  value  is not  unity,
consistent  with  the fact  that  the \pj  \  are  not uniform.   More
importantly,   this   choice    enables   us   to   obtain   converged
\widthidconvergencetest-\region \  width (\S\ref{section:results}) for
both     the     \lick    \     and     the    SDSS     configurations
(\figname\ref{fig:RegionWidth_NaD_vs_k_T0.70_Lick}                  and
\figname\ref{fig:RegionWidth_NaD_vs_k_T0.70_SDSS}, respectively).  The
atomic  lines,  such  as  \widthidconvergencetest,  have  well-defined
central wavelengths and widths.  They are therefore good standards for
checking  various identification  approaches.   For \pjsumthreshold  =
\chosenpjsumthreshold \ $\pm$  \chosenpjsumthresholdsd \ also give the
width convergence  in the SDSS configuration.  The  generality of this
convergence for arbitrary spectra is however not yet established.

A \roi  \ is fully indicated  by the left and  right wavelength bounds
\roiwave.  Once the  \wavec \ and the \roiwave  \ are determined, they
are labeled as  a ``\name'' and all of the  involved pixels are masked
out  in the  next  \wavec \  and  \roi \  selection.   The process  is
repeated until  either no  more pixels are  available, or  the desired
number of \region{s} is reached.  We then use
\begin{equation}
\mathrm{\pjsumname} = \pjsumregion = \sum_{\lambda = \roiws}^{\roiwe} p_{\lambda}
\end{equation}

\noindent 
as the measure for the information contained in a \region.

\subsection{Relation of \curonly \ to \pca}

 \pca  \ has  become a  standard technique  in spectral  analyses.  In
 extragalactic studies, it has been  applied to remove sky from galaxy
 spectra \citep{2005MNRAS.358.1083W},  to understand the  diversity in
 galaxies
 \citep[e.g.,][]{1995AJ....110.1071C,2004AJ....128..585Y,2012MNRAS.420.1217D}
 and                                                            quasars
 \citep[e.g.,][]{1992ApJ...398..476F,2003ApJ...586...52S,2004AJ....128.2603Y},
 to  separate  host galaxy  contribution  from  broadline AGN  spectra
 \citep{2006AJ....131...84V}, and  to find supernovae  in large galaxy
 spectral samples \citep{2003ApJ...599L..33M,2011ApJ...731...42K}, and
 many more studies.  These works  used the fact that \pca \ compresses
 the data in the {\it \idspace}, leaving the number of wavelength bins
 in each eigenspectrum  unchanged from that of the  input spectra. The
 \cur \ provides us  a new way to compress the data,  so that even the
      {\it \varspace}  can be compressed. The variable  in the current
      context is the wavelength.

\subsection{Spectrum Cutout Analysis} \label{section:cutout}

The next step is to compare the \region{s} quantitatively.  We prepare
spectral segments  of each \region \  that are {\it cut  out} from the
model spectra.   The set of  cutouts of each  \region \ form  a column
subspace  (in  $\realspace^{\anrow}$)  of  the original  $\anrow  \times
\ancol$  data matrix $A$  \citep[e.g.,][]{1988Strang}.  The  number of
cutout  pixels   is  less   than  the  number   of  spectra   in  this
work. Therefore, the rank of the subspace, at maximum, is equal to the
number of  pixels of that \region.   Its actual value  would depend on
the dimension  of the subspace  in question.  Three  subspace measures
are used: (I) the cosine of the angle between two subspaces as defined
by  \citet{2005GunawanNeswanSetyaBudhi},   \costheta.   (II)  The  dot
product between the first  left-singular vectors of the two subspaces,
\costhetafirstpc.  (III)  The Pearson correlation  coefficient between
the  integrated  fluxes   of  the  two  \region{s}\footnote{While  the
  subspaces spanned by the cutouts -- the spectra of narrow wavelength
  range  --  are  expected  to  be  fairly  linear  individually,  the
  correlation among the  subspaces may not be the  case.  In this work
  (\S\ref{section:importance},
  \figname\ref{fig:ar_AGE_orig_region1_JHU_k25.ecoeff_eigen1_Z__allbins}),
  we will show a~posteriori  that the Pearson correlation coefficient,
  a good  measure for linear correlation,  is unlikely to  be the best
  for probing the relation among \region{s}.}, \pearsoncoeff.

The  subspace measure  (I) is  calculated as  follows.  For  the $i$th
\region \  with number of  wavelength bins equal to  $\cutoutncol$, we
cut out from the original  data matrix $A$ the corresponding submatrix
$A^{(i)}$, size $\anrow \times \cutoutncol$.  A \svd \ is performed on
the  submatrix, $A^{(i)}  = U^{(i)}  \Sigma^{(i)}  {V^{T}}^{(i)}$. The
angle between the $i$th and the $j$th \region{s}, with $1 \leq n^{(i)}
\leq   n^{(j)}$,   is   calculated   by  applying   the   formula   in
\citet{2005GunawanNeswanSetyaBudhi}
\begin{equation}
\costhetamath = \sqrt{det(M^T M)} \ ,
\end{equation}

\noindent
where we put the matrix  $M = {U^{(j)}}^T {U^{(i)}}$. The symbol $det$
denotes the  determinant, and  $T$ the transpose,  of a  matrix.  They
have  shown that  \costheta \  is proportional  to the  volume  of the
parallelepiped spanned by  the projection of the basis  vectors of the
lower-dimensional  subspace on  the higher-dimensional  subspace.  The
one dimensional (1D) case is helpful  for us to grasp the picture: the
matrix  product $M^T M$  becomes the  dot product  between the  two 1D
basis vectors.

From  the \svd  \ of  the  \region{s} we  can also  calculate the  dot
product between  the first left-singular  vectors of any  two regions,
i.e., the subspace measure (II).  Indeed, any order of singular vector
can be  considered, but we pick  the first vector  because it captures
the maximum sample variance.

\subsection{Parameter Sensitivity of Regions}

To associate physical significance with the \region{s}, we explore the
correlation between a  chosen parameterization of a \region  \ and the
physical  parameters defining the  model.  We  start by  examining the
Pearson correlation  coefficient formula\footnote{Here we  examine the
  Pearson  correlation coefficient  from a  general perspective, in  a
  different  context  from  \S\ref{section:cutout}.}, the  correlation
between two variables $X$ and  $Y$ (each has $m$ components, $m$ being
the  number  of spectra/objects)  is  equivalent  to  the dot  product
between  two  vectors   in  the  \idspace  \  (each   vector  has  $m$
components).  We hereafter use ``the correlation between two subspaces
(in  the  \varspace)''  and  ``the  angle between  two  subspaces  (in
the \idspace)'' interchangeably.

This rather simple observation  is highly instrumental for quantifying
the parameter sensitivity  of a \region.  If the  subspace formed by a
\region \ is correlated with a  parameter, then the two vectors -- one
is  a chosen  parameterization of  the  subspace, and  another is  the
parameter  itself (vector  \parvector),  both have  $m$ components  --
should be  {\it parallel} to each  other in the  \idspace.  The vector
\parvector \ is either stellar  age or metallicity.  The angle between
the two vectors can be represented  by their dot product.  We pick two
parameterizations  for a  \region, namely,  the first  and  the second
left-singular   vectors    (vectors   \leftsingularvectorone   \   and
\leftsingularvectortwo), as most of the information is encapsulated by
the lowest-order modes.

\section{Results}\label{section:results}

\subsection{Informative Regions: \lick \ Configuration}

\begin{landscape}
\begin{figure}
  \plotone{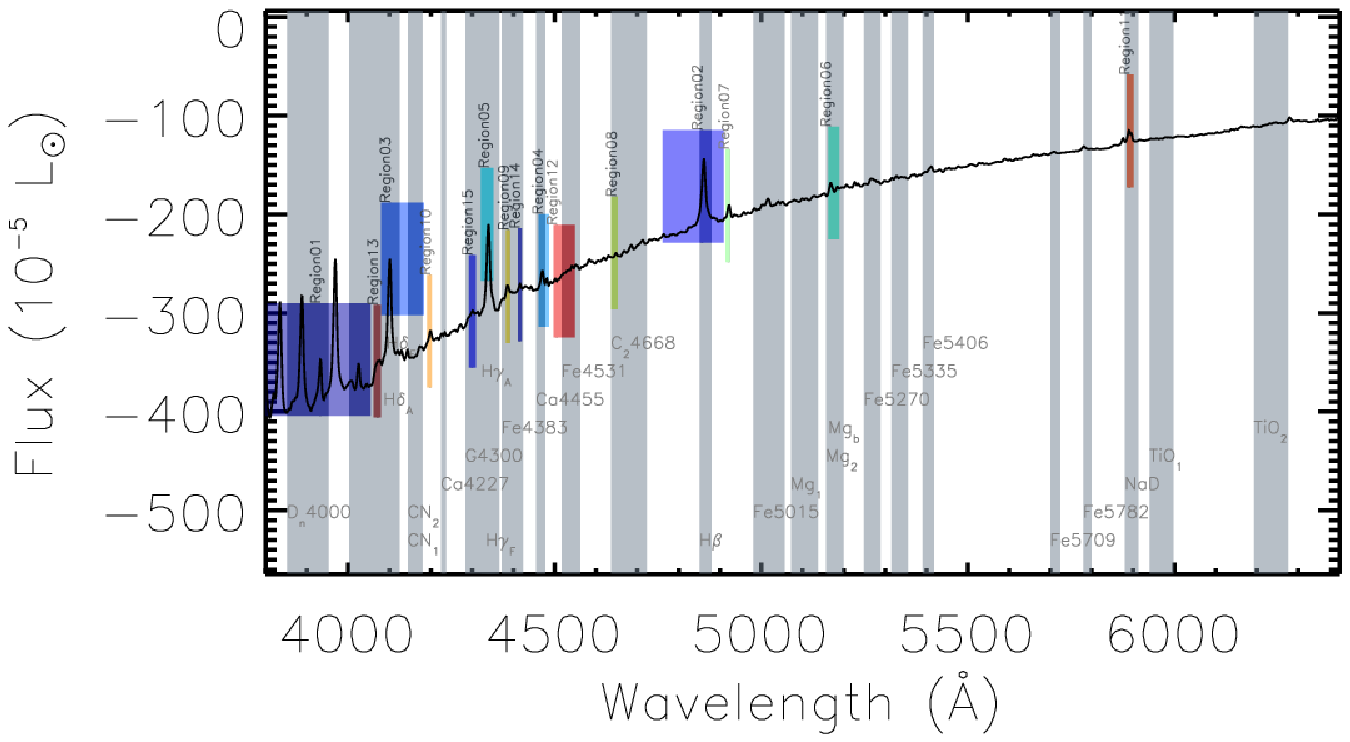}
\caption{The   first    \firstnlickconf   \   informative   wavelength
  \region{s},   identified  for   wavelengths  \lickconfwrange   \  at
  \lickconfres    \     spectral    resolution,    or     the    \lick
  \ configuration.  The gray  bars illustrate the  \extendedlick.  The
  first   three   regions   comprises   \lickconffirstimportantregion,
  \lickconfsecondimportantregion  \ and \lickconfthirdimportantregion.
  The wavelengths in the  vicinity of \lickconfunimportantregion \ are
  not selected, in qualitative  agreement with the \extendedlick.  The
  background is one of the  mean-subtracted spectra in the model.  The
  bottom of a \region \ label marks the \wavec \ of that region.  }
\label{fig:importantregions_k10_firstn20.transp}
\end{figure}
\end{landscape}

\begin{figure}
\plotone{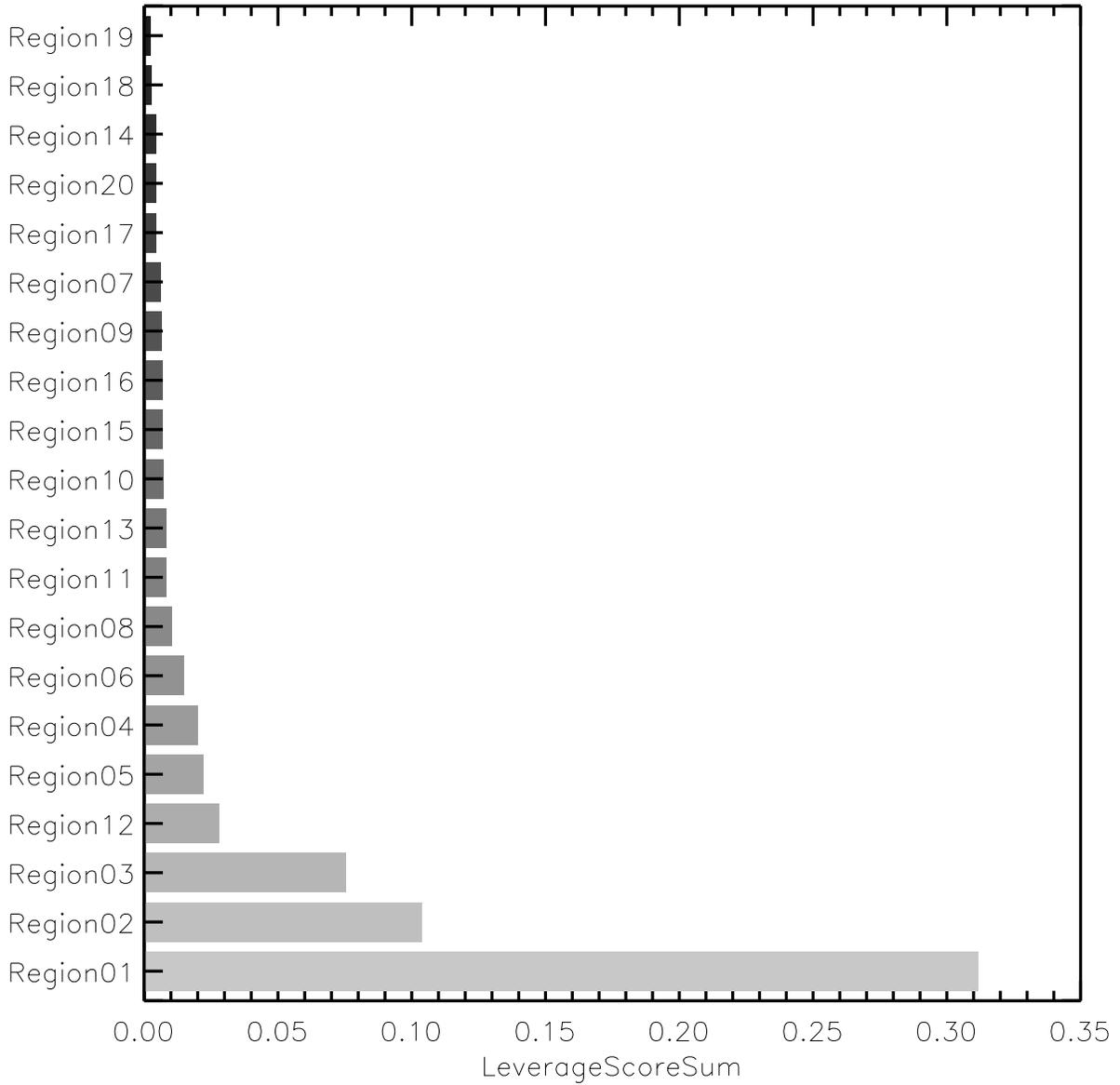}
\caption{The \pjsumname  \ of the  identified \region{s} in  the \lick
  \   configuration,    sorted   by   the    amplitude.    The   first
  \lickconfminregions \ \region{s} are most informative.}
\label{fig:pjsum_barplot_JHU_acc_k10.sorted}
\end{figure}

The  identified wavelength  regions using  our approach  are  shown in
\figname\ref{fig:importantregions_k10_firstn20.transp}.             The
corresponding  \wavec, $\roiws,  \roiwe$, as  well as  the overlapping
\extendedlick,  are  given  in \tabname\ref{tab:lickconf}.   From  the
width              convergence             of             \nad-\region
\ (\figname\ref{fig:pjsum_barplot_JHU_acc_k10.sorted}),  $k$ is set to
be \lickconfkpick.   The following \region{s} are found,  in the order
of importance:

\begin{itemize}
\item 1st: Comprises  the \lickconffirstimportantregion. The \pjsumname \  is a factor of a few or more larger than those of the other \region{s}.
\item 2nd: Comprises the  \lickconfsecondimportantregion \ line.
\item 3rd: Comprises   the \lickconfthirdimportantregion \ line. The \pjsumname \  is comparable to that of the \lickconfsecondimportantregion-region.
\item 4th: Comprises the \lickconffourthimportantregion, which belongs to the ``Fe-like indices'' family \citep{1998ApJS..116....1T}. 
\item 5th: Comprises the \gband. The \gband \ is primarily arisen from
  CH molecules and their energy  levels are therefore many. Not too surprisingly, it 
  is blended with \hgamma \ to form a single \region.  The \gband \ and  the \hgamma \ indices
  are also  not disjoint in wavelength in the \extendedlick \ definition (\tabname\ref{tab:lick}).
\item Higher orders: Certain  \extendedlick \  are recovered in
 the higher-order \region{s}.  
\end{itemize}

\begin{figure}
\plotone{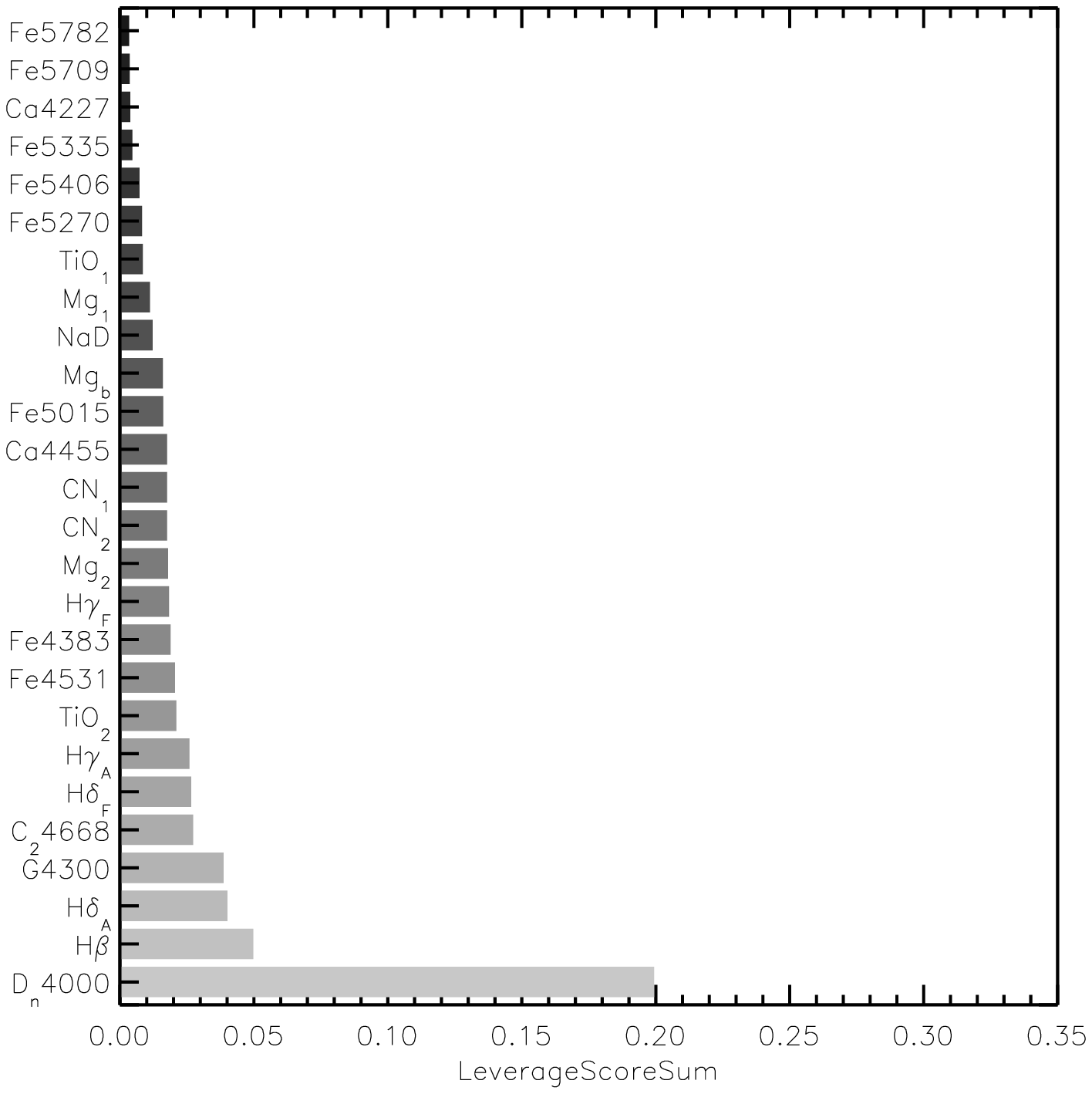}
\caption{The  \pjsumname  \  of   the  \extendedlick,  sorted  by  the
  amplitude.  We  note that  there was no  importance ordering  in the
  \extendedlick \ originally.   The indices \lickmostinformative \ are
  found to be most informative.}
\label{fig:pjsum_barplot_LICK_acc_k10.sorted}
\end{figure}

The    \pjsumname    \    of    our    indices    are    plotted    in
\figname\ref{fig:pjsum_barplot_JHU_acc_k10.sorted}.      The     first
\lickconfminregions   \  \region{s}  are   most  informative.    As  a
comparison, we  also calculate the \pjsumname \  of the \extendedlick,
shown  in  \figname\ref{fig:pjsum_barplot_LICK_acc_k10.sorted}.  There
was no importance ordering in  the \extendedlick \ originally but they
are   sorted    here   nonetheless.    The    widely   used   indices,
\lickmostinformative, are found indeed to be most informative.

We   also   see   that    the   wavelengths   in   the   vicinity   of
\lickconfunimportantregion   \  are   not  selected,   in  qualitative
agreement              with              the             \extendedlick
\    (\figname\ref{fig:importantregions_k10_firstn20.transp}).     The
wavelengths may be insensitive to stellar age nor stellar metallicity.

\subsection{Informative Regions: SDSS Configuration}

\begin{landscape}
\begin{figure}
\plotone{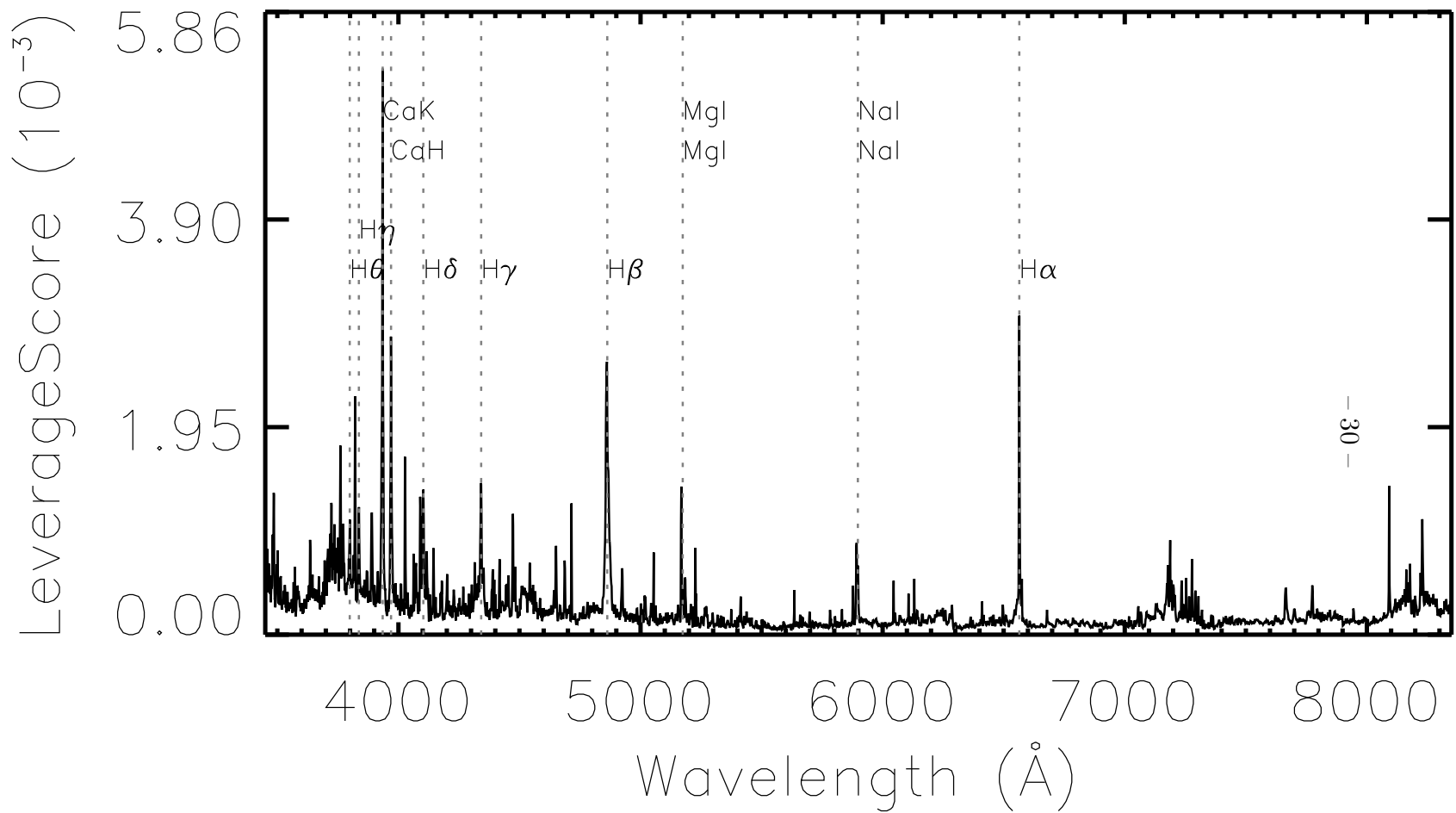}
\caption{The  \pjname  \  measures  the  information  contained  in  a
  wavelength of a set of  spectra.  In the SDSS  configuration.  Some
  absorption lines are marked.}
\label{fig:pj_25_25}
\end{figure}
\end{landscape}

\begin{landscape}
\begin{figure}
  \plotone{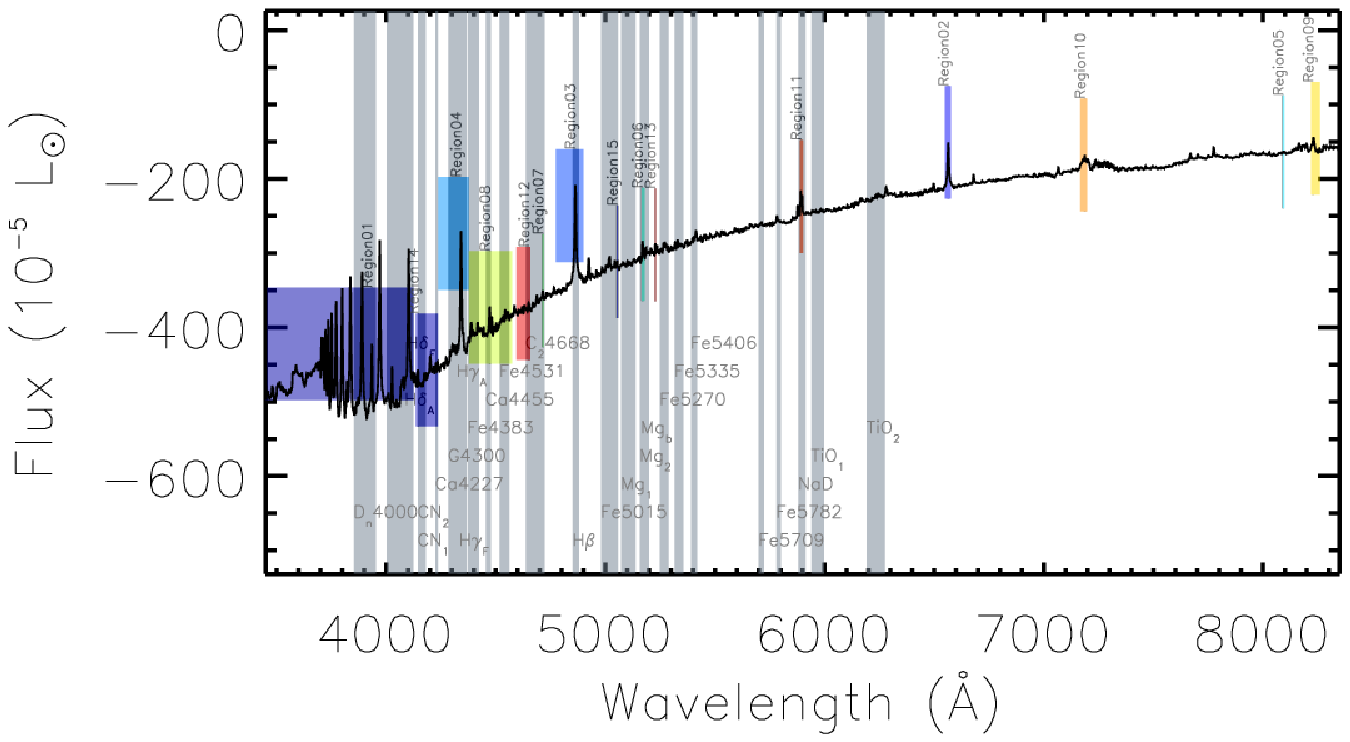}
\caption{The first \firstnsdssconf \ informative wavelength \region{s}
  identified  for  the wavelength  range  of  \sdssconfwrange  \ at  a
  spectral resolution of \sdssconfres, or the SDSS configuration.  The
  gray  bars  illustrate  the  \extendedlick.  Similar  to  the  \lick
  \ configuration, the \sdssconffirstimportantregion  \ is found to be
  most  informative; followed  by  the \sdssconfsecondimportantregion;
  the                \sdssconfthirdimportantregion;                the
  \sdssconffourthimportantregion.   The  background   is  one  of  the
  mean-subtracted spectra in the model.}
\label{fig:importantregions_k25_firstn20.transp}
\end{figure}
\end{landscape}

To  identify informative  \region{s}  in the  SDSS configuration,  the
first step is to determine the appropriate \neigen \ value. Because of
the higher resolution,  the rank of the data matrix  $A$ can be larger
than  the \lick  \ case  and the  appropriate \neigen  \ value  can be
different.  Examine  again the \nad-\region  \ width as a  function of
\neigen   \   (\figname\ref{fig:RegionWidth_NaD_vs_k_T0.70_SDSS}),  it
converges  to  \sdssconfwidthnad \  with  increasing  \neigen. We  use
\neigen  \ =  \sdssconfkpick \  in the  \region \  identification. The
\pjname   \   as   a    function   of   wavelength   is   plotted   in
\figname\ref{fig:pj_25_25},   with  high   amplitude   seen  in   some
absorption lines, and in  the vicinity of the \fourthousandbreak.  The
most       informative      \region{s}       are       plotted      in
\figname\ref{fig:importantregions_k25_firstn20.transp}.   They are, in
the order of importance:

\begin{itemize}
\item 1st: Comprises the \sdssconffirstimportantregion. In the \extendedlick, as we noted earlier, the \dn{R} and \hdeltaa \ overlap in wavelength. So it is not entirely surprising that they form a single \region. For example, there are likely many metal lines present in-between the \dn{L} and \dn{R}, so that the \dn{L}, \dn{R} and the \hdelta \ form one single \region.
\item 2nd: Comprises the \sdssconfsecondimportantregion~\citep{1998ApJS..116....1T}. This \region \ appears to be more sensitive to stellar metallicity than the most informative \region \ (the \sdssconffirstimportantregion), concluded from the dot product between the 2nd left-singular  and  stellar metallicity vectors (the last column of \tabname\ref{tab:sdssconf}). Unfortunately, the \pjsumname \ is a factor of \sdssconffirstsecondpjsumratio \ smaller, meaning this \region \ is less informative. We have yet connected the two different quantities -- the \pjsumname \ and the dot product -- to form a single measure for parameter sensitivity, which is beyond the scope of this work.
\item 3rd: Comprises the \sdssconfthirdimportantregion.
\item 4th: Comprises the \sdssconffourthimportantregion. The \pjsumname \ amplitudes are comparable in the \sdssconfpjsumcomparableregions \ most informative \region{s}, about \sdssconfpjsumcomparableamp.
\item Higher orders: The \sdssconfhalpharegionorder \ most informative \region \ comprises the \sdssconfhalphaimportantregion \ (\tabname\ref{tab:sdssconfcont}), which is detectable in most SDSS galaxies but falls outside of the \lick \ spectrograph wavelength range. 
\end{itemize}

\begin{landscape}
\begin{figure}
\plotone{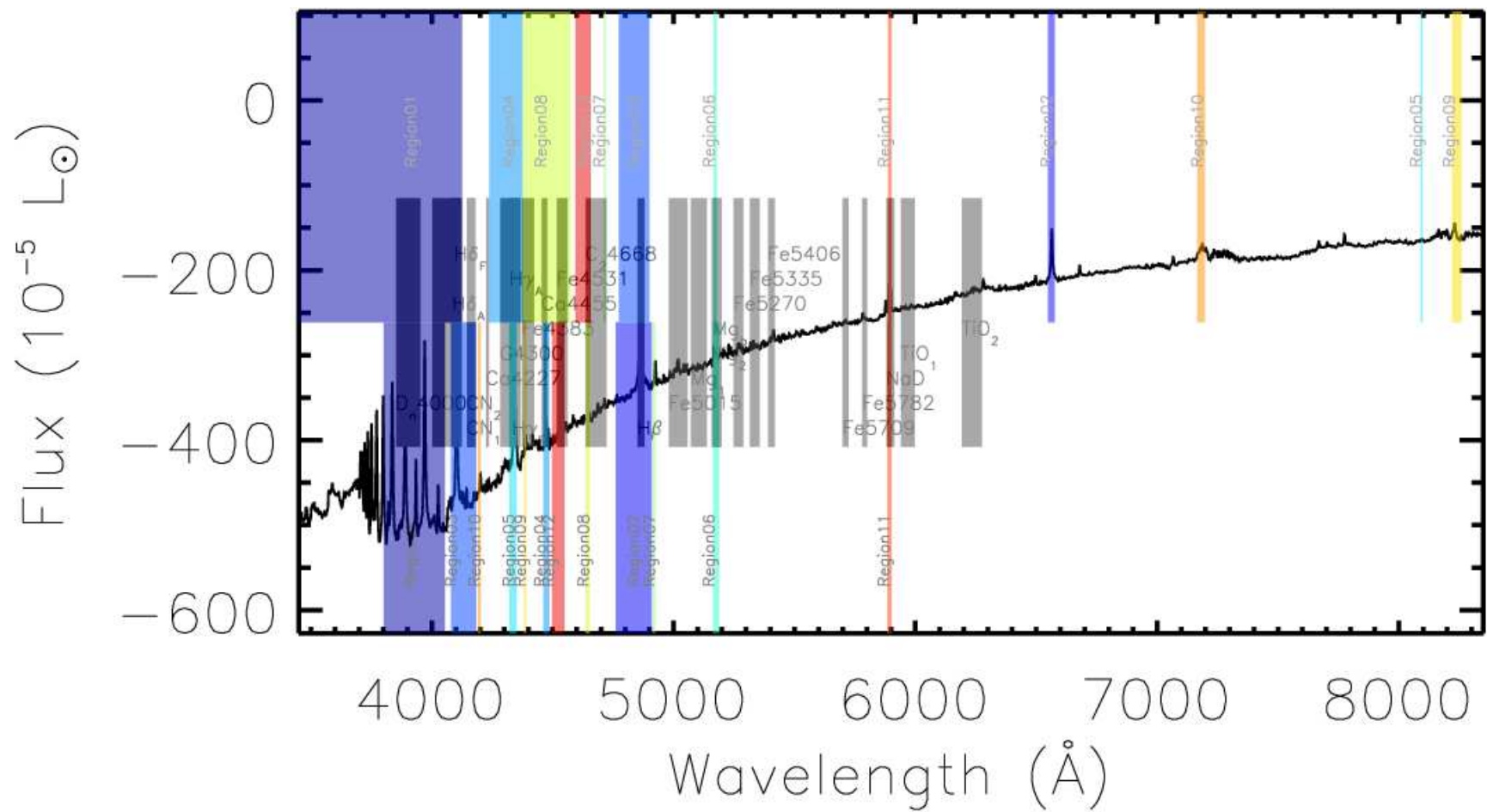}
\caption{Comparison   of  the   first  \firstncompare   \  informative
  \region{s}  that are  identified in  the \lick  \ (bottom)  and SDSS
  (top) configurations. The gray bars show the \extendedlick.}
\label{fig:importantregions_twosets.transp}
\end{figure}
\end{landscape}

 The comparison  of the  \region{s} between the  \lick \ and  the SDSS
 configurations              is             illustrated             in
 \figname\ref{fig:importantregions_twosets.transp}.   For  both  sets,
 the \breakfourthousand,  \hbeta, and  \hdelta \ \region{s}  are among
 the  top three most  informative.  We  however expect  the identified
 \region{s} to  depend on the  spectral resolution and  the wavelength
 coverage.   To  demonstrate  the wavelength-coverage  dependence,  an
 extra \region \ identification is  carried out where we use the \lick
 \ spectral resolution and pixel size but the SDSS wavelength coverage
 (not shown).  While  the \breakfourthousand \ remains to  be the most
 informative \region \ identified, its  \roi \ is diffferent from that
 obtained in  the \lick \ configuration,  and is more  similar to that
 obtained in the SDSS configuration.  One possibility is that the true
 \roi \ of the \breakfourthousand \ exceeds the shortest wavelength of
 the \lick \ configuration.

 The  dependence  of the  \region{s}  on  the  spectral resolution  is
 difficult to  be quantified generally.   When the resolution  is low,
 blended features  cannot be resolved.   The identified \region  \ can
 become broader than  the true wavelength bound, where  the latter may
 be  identifiable   only  in  higher-resolution   spectra.   When  the
 resolution is high,  the identified \region{s} may not  be optimal in
 terms of  studying lower-resolution spectra,  or they may not  be the
 state-of-the-art  for  future surveys.   To  avoid these  complicated
 situations,  it  is desirable  to  treat  the identification  method,
 instead of  the identified  \region{s}, to be  general. As  such, the
 method can be  applied to any set of spectra  of the user's interest.

\subsection{Relation Among Regions}

\begin{figure}
\plotone{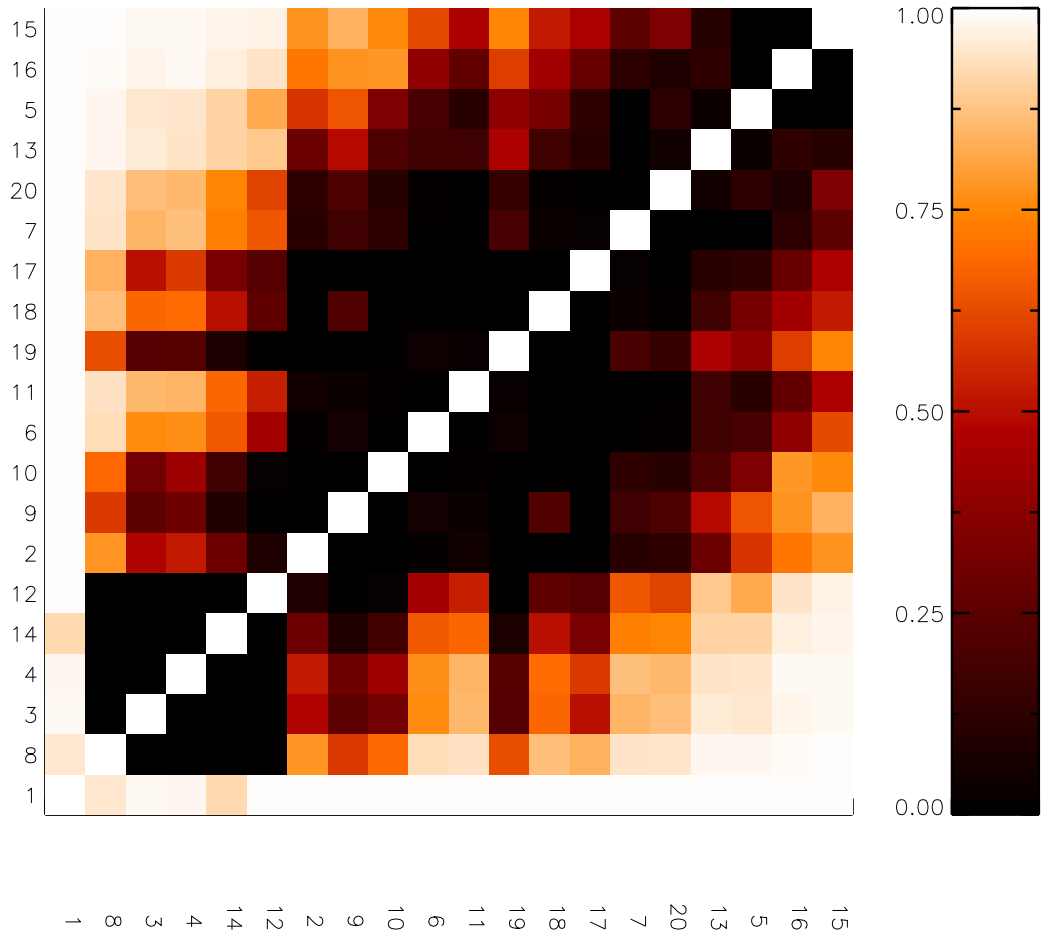}
\caption{The  \cossqtheta  \  between  the subspaces  spanned  by  the
  spectrum  cutouts   of  our  identified  \region{s},   in  the  SDSS
  configuration.   The line  indices  are sorted  by their  \pjsumname.    All   regions   are   almost   parallel   to   the
  \sdssconffirstimportantregionname.}
\label{fig:cutout_JHU_k25_sorted_costheta}
\end{figure}

\begin{figure}
\plotone{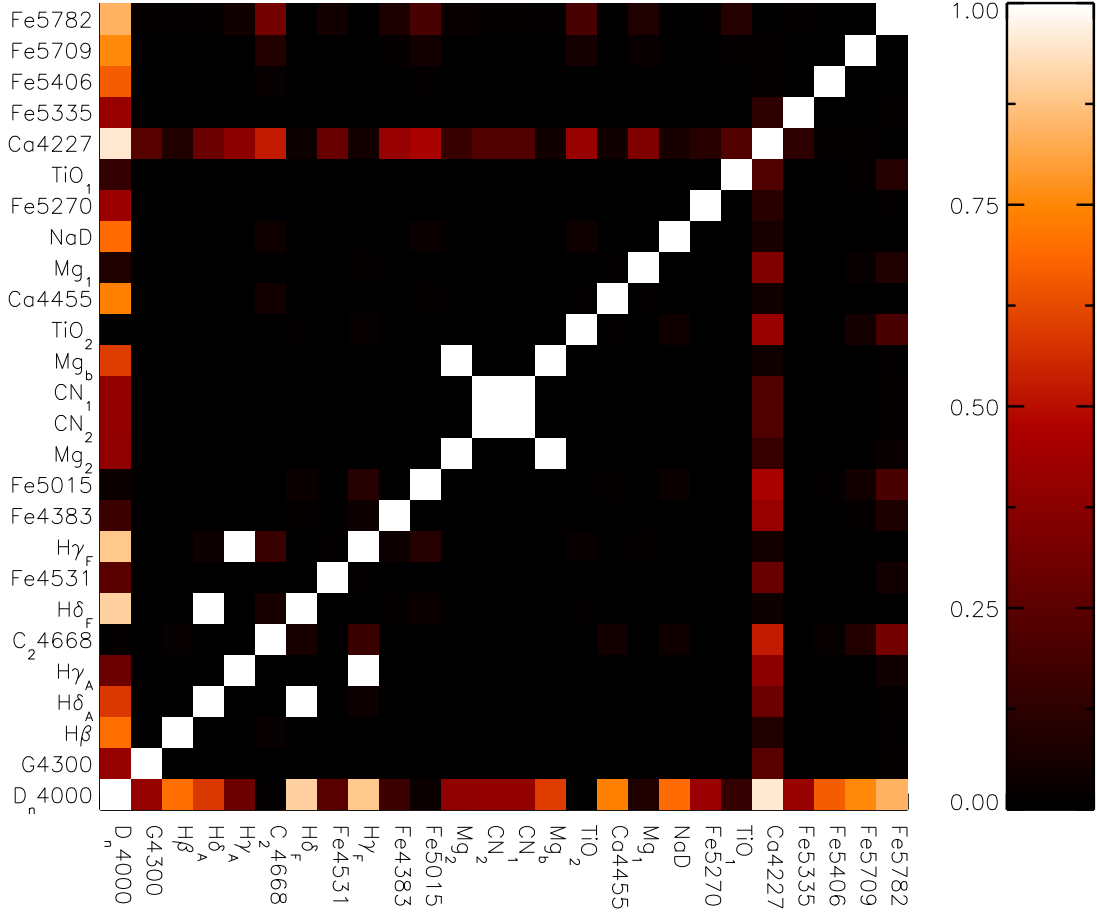}
\caption{The  \cossqtheta  \  between  the subspaces  spanned  by  the
  spectrum cutouts  of the  \extendedlick, in the  SDSS configuration.
  The  line  indices  are  sorted  by their  \pjsumname.
  Compared          with           the          \region{s}          in
  \figname\ref{fig:cutout_JHU_k25_sorted_costheta},   we  found  fewer
  indices to  be parallel to  the \fourthousandbreak. There is  also a
  marked                        difference          in  the number  of
  orthogonal index pairs.}
\label{fig:cutout_LICK_k25_sorted_costheta}
\end{figure}

The angle between the subspaces spanned by any two \region{s} is shown
in  \figname\ref{fig:cutout_JHU_k25_sorted_costheta},  in  \cossqtheta
\  \citep{2005GunawanNeswanSetyaBudhi}.  Those  for  the \extendedlick
\ are shown in \figname\ref{fig:cutout_LICK_k25_sorted_costheta}.  All
of   the    higher-order   \region{s}   are    correlated   with   the
\sdssconfminregions, suggesting  that the higher-order  \region{s} are
not substantially  different.  This is not  surprising considering the
\ssp{s} are defined by two parameters only, namely the stellar age and
metallicity, and  that the \breakfourthousand  \ is sensitive  to both
parameters.  Some of the \extendedlick  \ are also correlated with the
\dn     \    index,     but    not     as    many     as     in    the
\sdssconffirstimportantregionname,   nor  as   high   the  correlation
amplitudes.  This  result shows that \sdssconffirstimportantregionname
\  is  more  ``complete''  than   the  \dn,  in  the  sense  that  the
corresponding subspace encapsulates most of the data directions.

The other  pronounced difference between  the \extendedlick \  and the
\region{s} is that, in the former, there are many more cases where two
indices  are orthogonal.   The \region  \ representation  is therefore
more  ``compact'', in  the  sense  that we  need  fewer \region{s}  to
encapsulate the various data directions.

\begin{landscape}
\begin{figure}
\plottwo{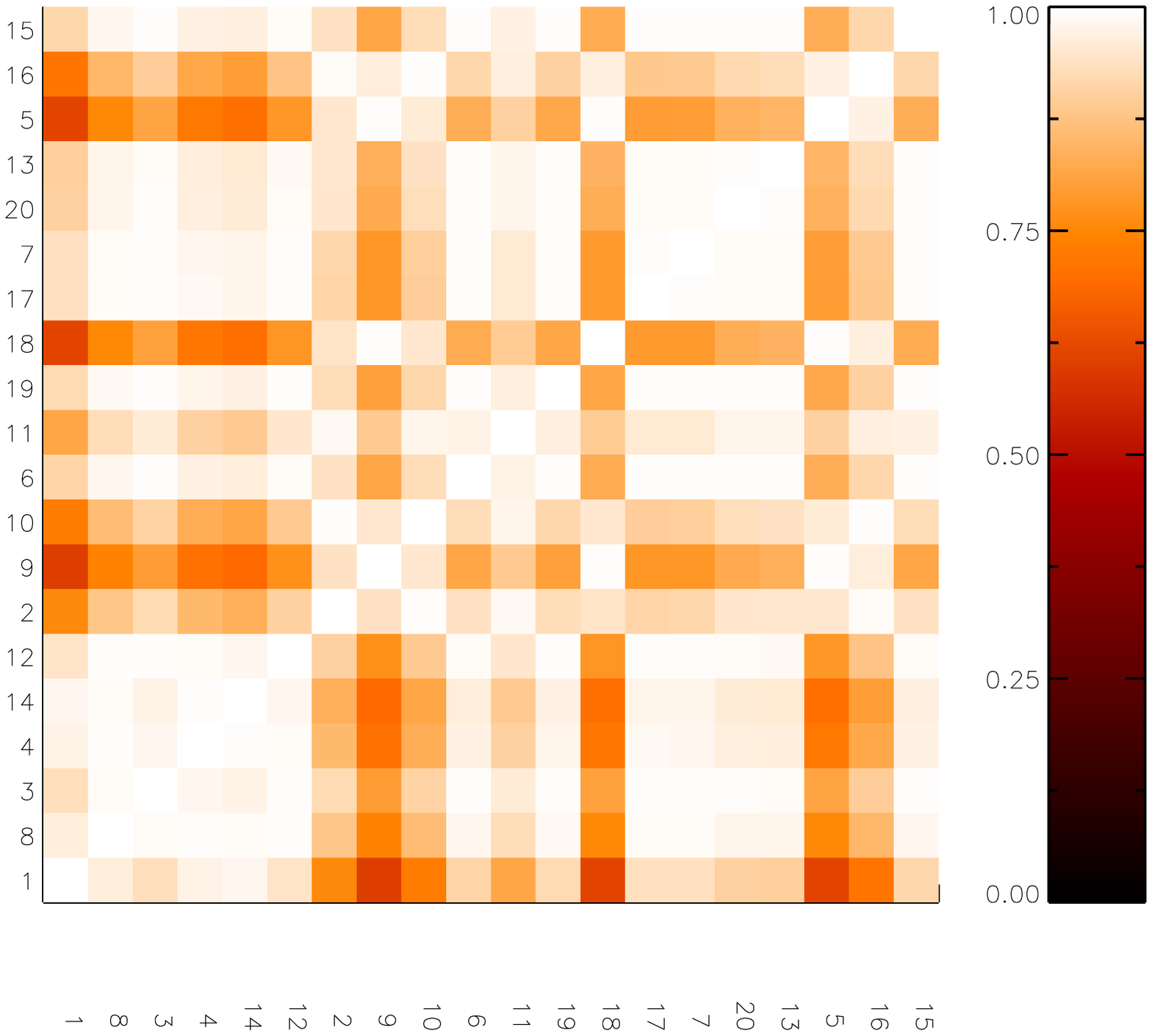}{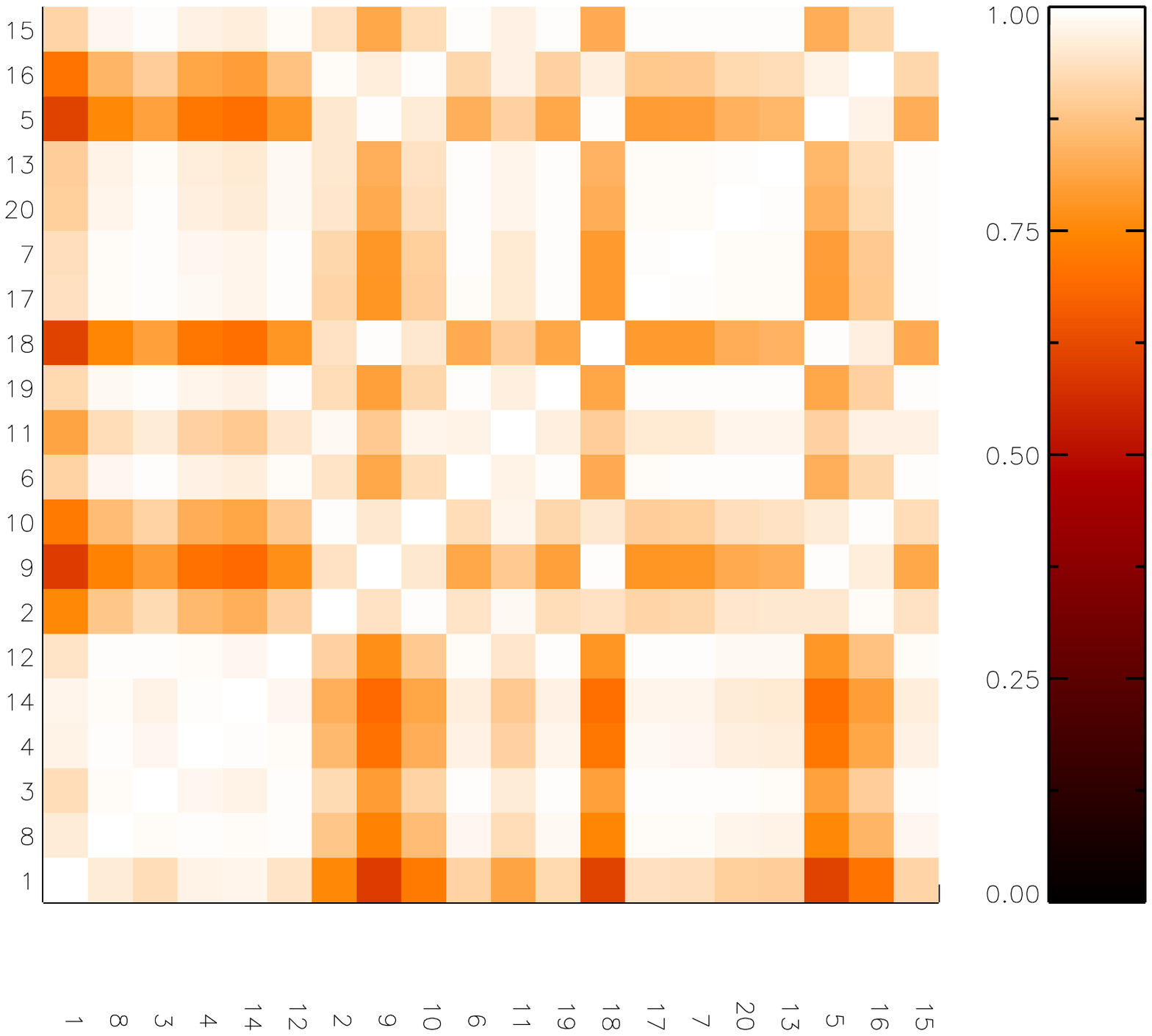}
\caption{(Left)   The    squared   Pearson   correlation   coefficient
  (\pearsoncoeffsq)  between  the  {\it  integrated flux} of the 
  identified \region{s}.  (Right) The  squared dot product between the
  first  left-singular vectors of  the {\it  spectral cutouts}  of the
  identified \region{s}  (\cossqthetafirstpc).  Both are  given in the
  SDSS configuration.  The diagonals show the auto-correlation of each
  \region. We found  both measures to be similar.}
\label{fig:cutout_JHU_k25_sorted_corrcoeff}
\end{figure}
\end{landscape}

The  squared\footnote{The   squared  Pearson  correlation  coefficient
  (\pearsoncoeffsq)  is used  because we  would like  to focus  on the
  amplitude  of the  correlation.  It  ranges  from 0  to 1,  0 if  no
  correlation,  1 if  100\% correlation.   A consequence  is  the same
  color coding  in the \pearsoncoeffsq  \ and \cossqtheta  \ figures.}
Pearson   correlation   coefficient   (\pearsoncoeffsq)  between   the
integrated flux of two \region{s}, and the squared dot product between
the first left-singular vectors of two subspaces (\cossqthetafirstpc),
are                shown                side-by-side                in
\figname\ref{fig:cutout_JHU_k25_sorted_corrcoeff}.   Large correlation
amplitude  is  seen  between  most \region \ pairs.         Using  the
integrated flux or  the first singular vector to  represent a subspace
hence convey less information about  that subspace than using a number
of  singular vectors.   This  result is  not  surprising, because  the
integrated  flux does not  fully describe  a \region,  or in  fact any
spectrum.  This  result  also justifies  the  use  of  the \pca  \  to
determine physical parameters  from a galaxy spectrum.  Interestingly,
from  \figname\ref{fig:cutout_JHU_k25_sorted_corrcoeff}  we also  find
both measures  to be similar.   A mathematical explanation  is however
not yet explored.

\subsection{Physical Significance of Regions}\label{section:importance}

 The parameter sensitivity  of the \region{s} \ are given  in the last
 few       columns       of       \tabname\ref{tab:lickconf}       and
 \tabname\ref{tab:sdssconf},  respectively for  the \lick  \ and  SDSS
 configurations.  All \region{s}  are sensitive to \firstpar  \ to the
 first order, and to either \firstpar  \ or \secondpar \ to the second
 order.  \citet[][]{2012ApJ...756..163S}  have also shown that  age is
 the  main parameter  driving the  variance  in the  spectra of  local
 galaxies.  We expect some \region{s} will be sensitive to the stellar
 metallicity to  the first order  if the input spectra  were continuum
 subtracted, taking out  the sample variance that is  primarily due to
 stellar age.   While this is  an interesting alternative to  the data
 centering (in the preprocessing,  \S\ref{section:data}), care must be
 taken to  propagate the uncertainty  of the estimated  continuum into
 the \region{s} identification.


We    then    perform    a    \pca    \   on    the    cutouts    from
\sdssconffirstimportantregionname  \  in  the SDSS  configuration  and
relate  the eigencoefficients  to  stellar age  and metallicity.   The
eigenspectra are  shown in \app\ref{appendix:espec}.   Focusing on the
lowest      orders       of      eigencoefficients,      shown      in
\figname\ref{fig:ar_AGE_orig_region1_JHU_k25.ecoeff_eigen1_Z__allbins},
the  first  eigencoefficient  correlates  well with  the  stellar  age
regardless of the stellar metallicity, and the second eigencoefficient
with  the stellar  metallicity {\it  after the  age is  known}.  These
results  agree  with those  obtained  from  the parameter  sensitivity
analysis   that    are   shown   in   the   last    few   columns   of
\tabname\ref{tab:sdssconf}.  The determination of stellar age from the
first  eigencoefficient appears  to work  best for  intermediate ages,
which is  not surprising considering  the optical spectra  of galaxies
dominated  by old  ($>$  a few~Gyr)  stellar  populations look  alike,
posing a well-known limitation on  the determination of stellar age in
galaxies from optical spectra.  We  also found a similar situation for
galaxies  dominated  by very  young  populations  ($<$ 10~Myr),  where
spectra of two different ages can show the same first-eigencoefficient
amplitude.  Interestingly, we  have to know the stellar  age before we
can  tell the stellar  metallicity, in  other words,  they have  to be
determined simultaneously, which is  a manifestation of the well-known
"age-metallicity                  degeneracy"                  problem
\citep[e.g.,][]{1973ApJ...179..731F,1978ApJ...220..434M,1989MNRAS.238..925S}
in  galaxy  parameter  estimation.    To  conclude,  we  can  use  the
\region{al}  eigencoefficients  to   determine  the  stellar  age  and
metallicity in early-type galaxies  which have solar abundance ratios,
no dust, and a single-burst  star formation history.  It remains to be
seen how the parameters which are relevant to late-type galaxies, such
as  the exponentially  decreasing  star formation  rate  and the  dust
extinction, depend on the eigencoefficients of a \region.

\section{Conclusions}\label{section:summary}

  We  identify  informative wavelength  \region{s}  in a  single-burst
  stellar     populations     model      by     using     the     \cur
  \  \citep{2008Drineasetal,2009MahoneyDrineas}.   The \region{s}  are
  objective.   They are  shown  to  be sensitive  to  stellar age  and
  metallicity.  The  \region{s} can be  used to determine  the stellar
  age  and  metallicity  in   early-type  galaxies  which  have  solar
  abundance  ratios,  no  dust,  and  a  single-burst  star  formation
  history.   The  \region \  identification  method  and the  subspace
  analysis  can  be  applied to  any  set  of  spectra of  the  user's
  interest,  so  that  we  eliminate  the need  for  a  common,  fixed
  resolution index system.



 We plan to extend this  analysis to late-type galaxies.  The presence
 of   emission  lines   pose   special  challenges   to  the   \region
 \  identification  on the  whole  spectra,  namely,  the continuum  +
 absorption  + emission spectra.   This speculation  is hinted  by the
 fact  that the  continuum-included  strong emission  lines cannot  be
 reconstructed with higher than \linereconacc \ accuracy on average by
 using      a       handful      of      lower-order      eigenspectra
 \citep[][]{2004AJ....128..585Y,2012arXiv1207.4374M}.       We     are
 investigating  this  question.   A   possible  approach  is  that  of
 \citet{2011AJ....141..133G}        who        parameterized       the
 continuum-subtracted   emission-line  EWs   through   a  handful   of
 eigenspectra.   To this  end, the  SDSS  galaxies will  be a  perfect
 dataset.  Because of the many  galaxy types, a large diversity in the
 emission lines and  the associated gas kinematics can  be studied.  A
 set of \region{s}, taken into account of both absorption and emission
 lines,  will give  a comprehensive  parametrization  to emission-line
 galaxy spectra.  The progress on understanding galaxy parameters have
 already        set        benchmarks        in       the        field
 \citep[e.g.,][]{2003MNRAS.341...33K,2005MNRAS.362...41G,2007MNRAS.381..543W,2012MNRAS.421..314C}.
 This work  provides a factor  of \datareduction \ reduction  over the
 original data.  Such a data compression will be crucial if we want to
 estimate   simultaneously  a  large   number  (\manypar)   of  galaxy
 parameters~\citep{2007IAUS..241..533Y,2010AJ....139..342Y}   in   the
 future,  such as stellar  age and  metallicity, dust  extinction, and
 high temporal resolution star formation history.

\section{Acknowledgments}

We thank  Andrew~Connolly, Haijun~Tian, Miguel~Angel~Aragon~Calvo, and
Guangtun~Zhu  for   useful  comments  and   discussions.   CWY  thanks
Scott~Trager for discussions on  galaxy spectra.  We thank the referee
for careful reading of the manuscript and useful suggestions.

This  research  is  partly  funded  by  the  Gordon  and  Betty  Moore
Foundation through Grant GBMF\#554.01 to the Johns Hopkins University.
IC and LD  acknowledge grant OTKA-103244.  This research  has made use
of data obtained from or  software provided by the US National Virtual
Observatory, which is sponsored by the National Science Foundation.

{}

\clearpage

\appendix
\section{\cur} \label{appendix:cur}

The  \cur   \  \citep{2008Drineasetal,2009MahoneyDrineas}  computes  a
low-rank approximation to an arbitrary matrix that marvels the optimal
low-rank  approximation  provided  by  the  truncated  Singular  Value
Decomposition  (\svd).  The  approximation, however,  is  expressed in
terms of  a small number of  \emph{actual} columns and/or  rows of the
input  data  matrix.  That  is,  it  captures  the dominant  modes  of
variation in  a data matrix  with a small  number of actual  (and thus
potentially interpretable)  columns and/or  rows, rather than  a small
number of  (in general non-interpretable)  eigencolumns and eigenrows.
The  \cur \  has been  used in  Genetics to  identify ``PCA-correlated
SNPs'',   basically   the   most   informative  columns   within   DNA
single-nucleotide polymorphism  (SNP) matrices~\citep{Paschou07b}; and
it has been central to recent work in developing randomized algorithms
for     the      low-rank     approximation     of      very     large
matrices~\citep{2011arXiv1104.5557M}.

The main  idea behind the  \cur \ is  to decompose a given  matrix $A$
into  matrices $C$  and $R$  which  consist of  respectively a  small
number  of actual  columns  and  rows of  $A$,  and a  low-dimensional
encoding matrix  $\mathbb{U}$, such that $A  \approx C\mathbb{U}R$ as
follows

\begin{equation}
\|A-C\mathbb{U}R\|_F \le (1+\epsilon)\|A-A_k\|_F ,
\label{eqn:cur}
\end{equation}

\noindent
where $A_k$ is the best rank-$k$ approximation to $A$, as given by the
truncated  \svd.   The  fractional   error  of  the  decomposition  is
$\epsilon$.  Subscript $F$ denotes the Frobenius norm of a matrix.  We
use  only the column  part of  the decomposition  in this  paper.  The
choice of  the columns is  critical: to obtain  low-rank approximation
bounds  of  the form  in  \eqnname\ref{eqn:cur},  one chooses  columns
randomly  according to  an  importance sampling  distribution that  is
proportional to the  \score{s} (of $A$, relative to  the best rank-$k$
approximation    to     $A$).     These    quantities,     given    in
\eqnname\ref{eqn:pj},  equal the diagonal  elements of  the projection
matrix onto the  span of the best rank-$k$  approximation to $A$. They
have been  used previously in  regression diagnostics as a  measure of
the  importance   or  influence  a   given  data  point  has   on  the
least-squares  fit~\citep{CH86}.   When  applied  to  low-rank  matrix
approximations, the \score{s} provide  a measure of how informative is
a  given  column  to  the  best rank-$k$  approximation  of  the  data
matrix. In our case, a column is a wavelength in the model spectra.



\clearpage

\section{Regional Eigenspectra} \label{appendix:espec}

\begin{figure}
\plotone{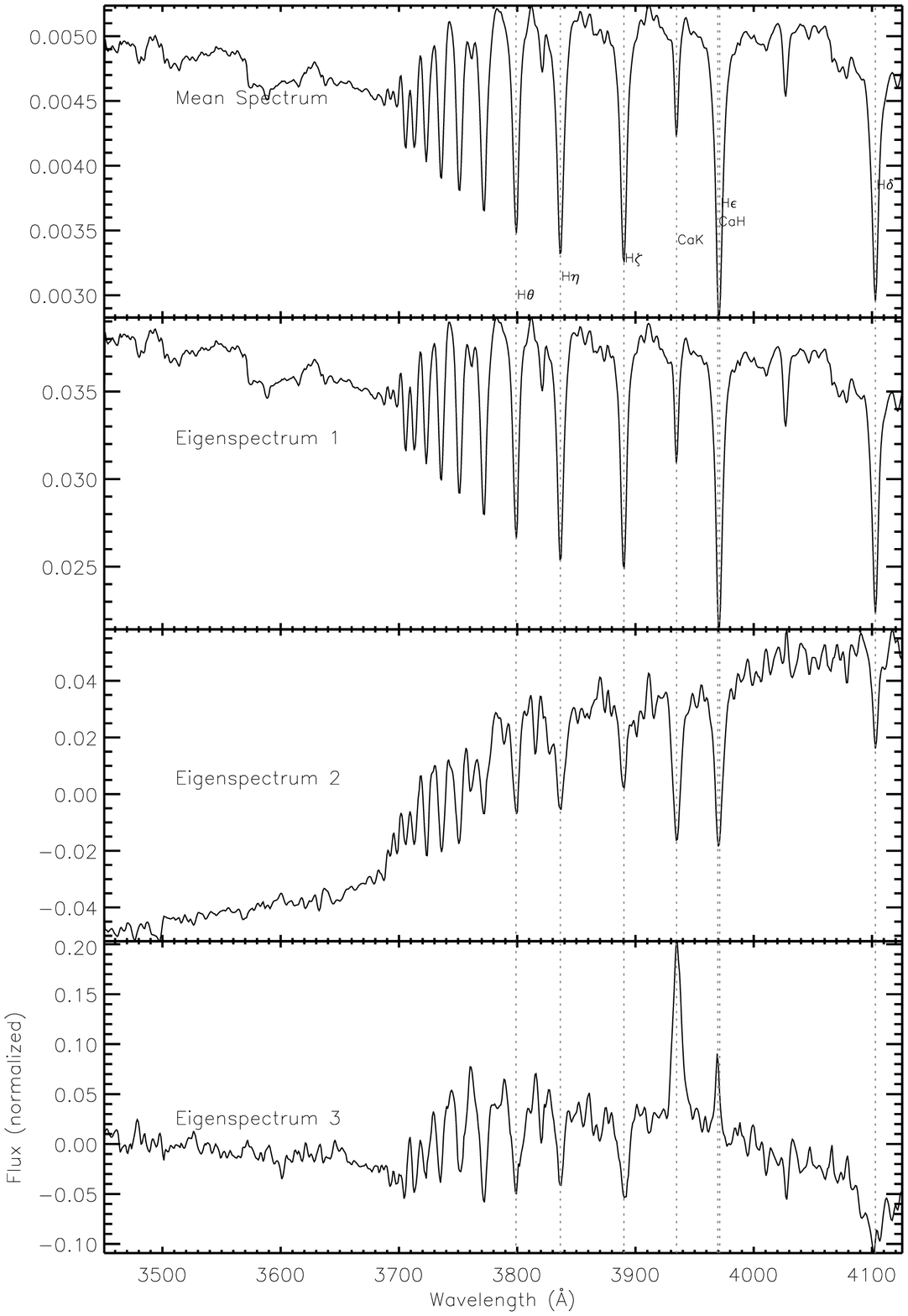}
\caption{The  mean  spectrum (top)  and  first  three eigenspectra  of
  \sdssconffirstimportantregionname \ in  the SDSS configuration.  The
  first  eigenspectrum modulates the  \firstespecshows \  strength (i.e.,
  the   \firstespec).    The   second  eigenspectrum   modulates   the
  \secondespecshows  \  strength  (i.e., the \secondespec).   The  third
  eigenspectrum shows an \thirdespec.}
\label{fig:espec_orig_region1_JHU_k25_mean_and_espec}
\end{figure}

\begin{landscape}
\begin{figure}
\plottwo{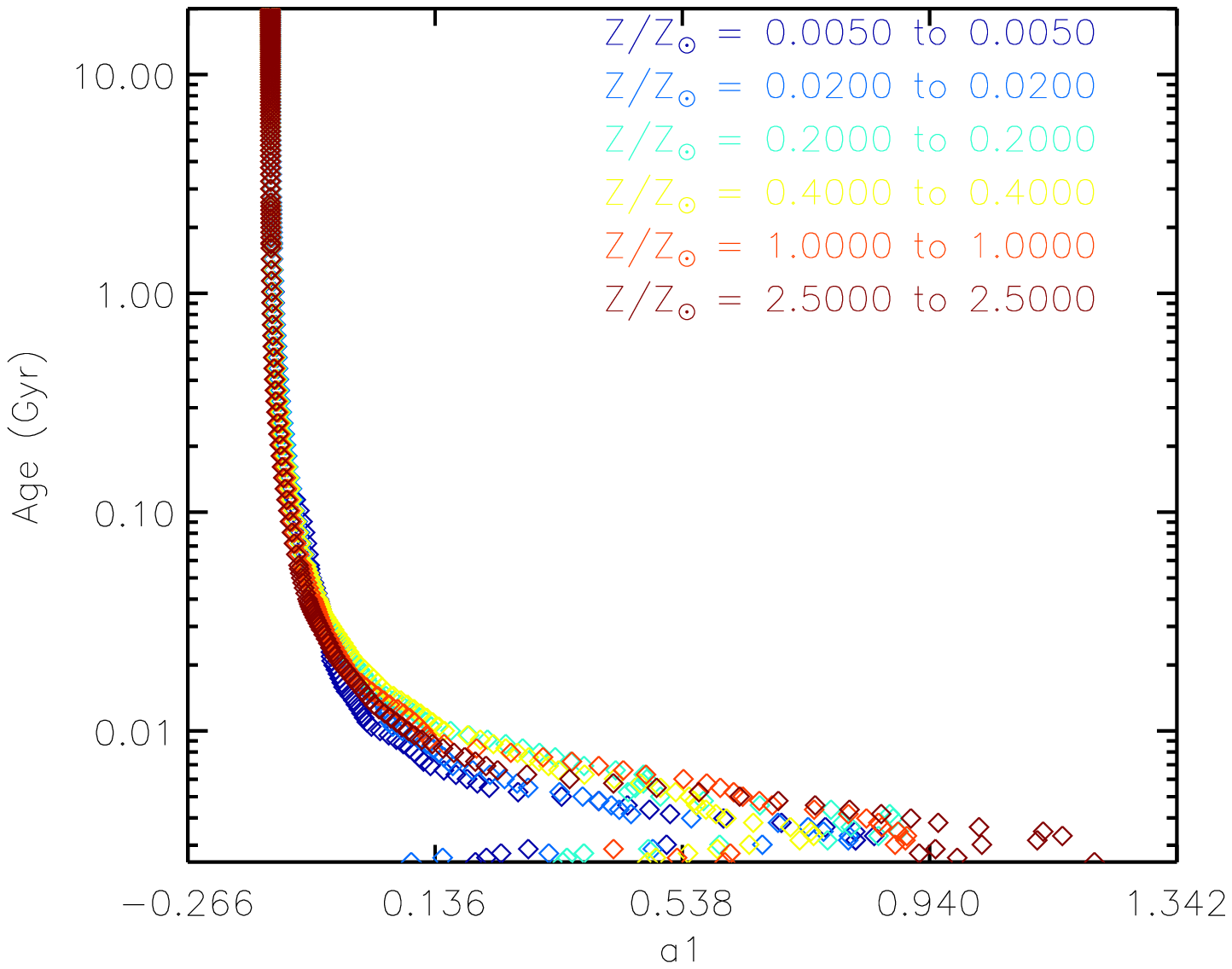}{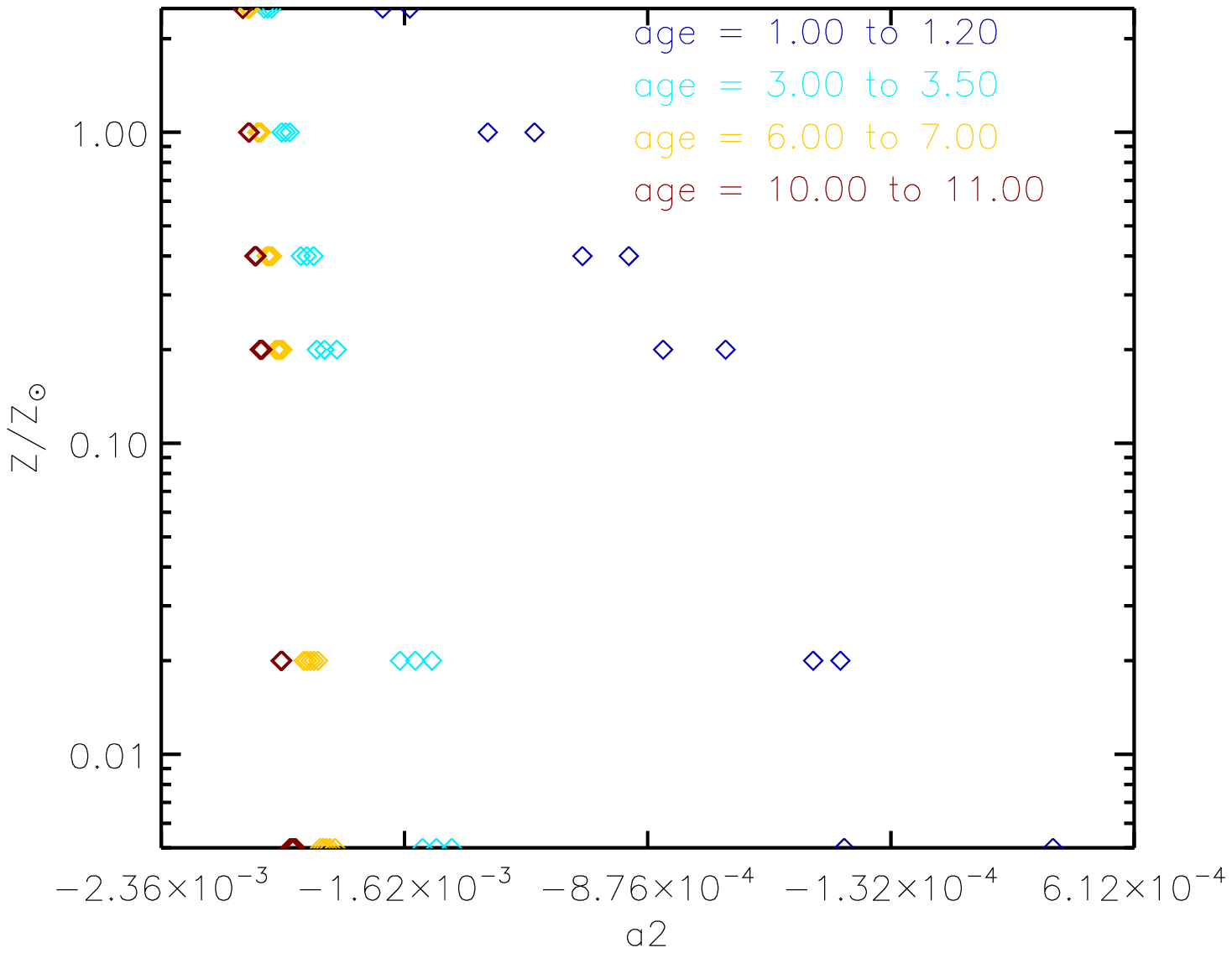}
\caption{(Left)  The  dependence of  stellar  age  on  the first  \pca
  \   eigencoefficient   of   \sdssconffirstimportantregionname,   for
  different stellar metallicities.   (Right) The dependence of stellar
  metallicity  on the  second \pca  \ eigencoefficient,  for different
  stellar ages, in the  SDSS configuration. Although the relations are
  monotonic  (except the  smallest ages  on  the left),  they are  not
  straightly  linear,  which  in  turn explains  the small correlation
  coefficient  amplitude listed  in  \tabname\ref{tab:sdssconf}.  This
  result  suggests that  other (nonlinear)  measures, rather  than the
  Pearson  correlation coefficient, may  better describe  the relation
  between the eigencoefficients and the physical parameters.}
\label{fig:ar_AGE_orig_region1_JHU_k25.ecoeff_eigen1_Z__allbins}
\end{figure}
\end{landscape}

The lowest orders  of the eigenspectra encapsulate most  of the sample
variance   in   the   full   optical  spectra   of   nearby   galaxies
\citep[][]{2004AJ....128..585Y},  forming a subspace  which lies  in a
higher-dimensional wavelength  space.  We  examine here the  first few
orders of  eigenspectra of \sdssconffirstimportantregionname  \ in the
SDSS configuration.  Together with the  mean spectrum of the model the
eigenspectra                are               plotted               in
\figname\ref{fig:espec_orig_region1_JHU_k25_mean_and_espec}.        The
third eigenspectrum shows an \thirdespec.  \citet{2007MNRAS.381..543W}
also  found that  the third  mode modulates  the Ca~K  \&  H strength,
though the  concerned model has an exponential  star formation history
with      recent      stellar      bursts.       We      see      from
\figname\ref{fig:espec_orig_region1_JHU_k25_mean_and_espec}  that  the
first    and   second    eigenspectra   modulate    respectively   the
\firstespecshows  \ and  the  \secondespecshows \  strengths.  On  the
other   hand,    we   know   from   the    parameter   dependence   in
\figname\ref{fig:ar_AGE_orig_region1_JHU_k25.ecoeff_eigen1_Z__allbins}
that first eigenspectrum of the single-burst stellar populations model
tells primarily the \firstespec.   The second eigenspectrum is related
to the \secondespec.  We therefore conclude the following.  The larger
the  first eigencoefficient  of a  galaxy spectrum,  the  stronger the
Balmer  absorptions, indicating the  stellar populations  are younger.
The   larger   the   second   eigencoefficient,   the   stronger   the
\breakfourthousand,  indicating  the  stellar populations  are  either
older and less metal rich, or younger and more metal rich.

\epsscale{0.8}

\begin{table}
\begin{center}
\caption{\pjsumname \ of \extendedlick, for wavelength range \lickconfwrange \ at \lickconfres \ spectral resolution.}
\begin{tabular}{llccccc}
\hline
Region & Reference & Unit & $\lambda_{S}$ & $\lambda_{E}$ & $\lambda_{E} - \lambda_{S}$ & $\sum p_{\lambda}$ \\
\hline
D$_{n}$(4000) & Balogh99 & N/A & [3851.1, 4001.1] & [3951.1, 4101.2] & [100.0, 100.0] & 0.199 \\
H$\beta$ & Trager98 & \AA & 4849.2 & 4878.0 &   28.8 & 0.050 \\
H$\delta_{A}$ & WortheyOttaviani97 & \AA & 4084.7 & 4123.4 &   38.8 & 0.040 \\
G4300 & Trager98 & \AA & 4282.6 & 4362.6 &   80.0 & 0.039 \\
C$_{2}$4668 & Trager98 & \AA & 4635.3 & 4721.6 &   86.3 & 0.027 \\
H$\delta_{F}$ & WortheyOttaviani97 & \AA & 4092.2 & 4113.4 &   21.3 & 0.027 \\
H$\gamma_{A}$ & WortheyOttaviani97 & \AA & 4321.0 & 4364.7 &   43.8 & 0.026 \\
TiO$_{2}$ & Trager98 & mag & 6191.3 & 6273.9 &   82.5 & 0.021 \\
Fe4531 & Trager98 & \AA & 4515.5 & 4560.5 &   45.0 & 0.021 \\
Fe4383 & Trager98 & \AA & 4370.4 & 4421.6 &   51.3 & 0.019 \\
H$\gamma_{F}$ & WortheyOttaviani97 & \AA & 4332.5 & 4353.5 &   21.0 & 0.018 \\
Mg$_{2}$ & Trager98 & mag & 5155.6 & 5198.1 &   42.5 & 0.018 \\
CN$_{2}$ & Trager98 & mag & 4143.3 & 4178.3 &   35.0 & 0.018 \\
CN$_{1}$ & Trager98 & mag & 4143.3 & 4178.3 &   35.0 & 0.018 \\
Ca4455 & Trager98 & \AA & 4453.4 & 4475.9 &   22.5 & 0.018 \\
Fe5015 & Trager98 & \AA & 4979.1 & 5055.4 &   76.3 & 0.016 \\
Mg$_{b}$ & Trager98 & \AA & 5161.6 & 5194.1 &   32.5 & 0.016 \\
NaD & Trager98 & \AA & 5878.5 & 5911.0 &   32.5 & 0.012 \\
Mg$_{1}$ & Trager98 & mag & 5070.5 & 5135.6 &   65.0 & 0.011 \\
TiO$_{1}$ & Trager98 & mag & 5938.3 & 5995.8 &   57.5 & 0.009 \\
Fe5270 & Trager98 & \AA & 5247.1 & 5287.1 &   40.0 & 0.008 \\
Fe5406 & Trager98 & \AA & 5389.0 & 5416.5 &   27.5 & 0.007 \\
Fe5335 & Trager98 & \AA & 5313.6 & 5353.6 &   40.0 & 0.005 \\
Ca4227 & Trager98 & \AA & 4223.4 & 4235.9 &   12.5 & 0.004 \\
Fe5709 & Trager98 & \AA & 5698.2 & 5722.0 &   23.8 & 0.004 \\
Fe5782 & Trager98 & \AA & 5778.2 & 5798.2 &   20.0 & 0.003 \\
\hline
\end{tabular}
\tablecomments{The \extendedlick \ defined in this work, in vacuum
  wavelengths. The indices are ordered by their \pjsumname, as listed in
  the last column. The $\roiws$  and $\roiwe$ are the index wavelength bounds taken from the references.}
\label{tab:lick}
\end{center}
\end{table}

\begin{sidewaystable}
{\footnotesize
\caption{Informative wavelength \region{s}, for wavelength range \lickconfwrange \ at \lickconfres \ spectral resolution.}
\begin{tabular}{ccccccccccrcr}
\hline
Region & $\lambda_{S}$ & $\lambda_{I}$ & $\lambda_{E}$ & $\lambda_{E} - \lambda_{S}$ & $\lambda_{I} - \lambda_{S}$ & \rotatebox{90}{Symmetric?}& $\sum p_{\lambda}$ & \lick \ Overlaps & \rotatebox{90}{1st mode}  & \rotatebox{90}{$u^{1} \cdot$ \parvector}  &\rotatebox{90}{2nd mode}  & \rotatebox{90}{$u^{2} \cdot$ \parvector}  \\
\hline
Region01 &  3801.2 &  3932.5 &  4053.8 &   252.5 &   131.2 & no &0.312 &  D$_{n}$(4000)L;D$_{n}$(4000)R; & AGE   & -0.395 & METAL & -0.135 \\
Region02 &  4761.2 &  4856.2 &  4908.8 &   147.5 &    95.0 & no &0.104 &  H$\beta$; & AGE   & -0.393 & METAL & -0.163 \\
Region03 &  4080.0 &  4101.2 &  4183.8 &   103.8 &    21.2 & no &0.075 &  CN$_{1}$;CN$_{2}$;H$\delta_{A}$;H$\delta_{F}$;D$_{n}$(4000)R; & AGE   & -0.396 & AGE   &  0.208 \\
\hline
Region12 &  4497.5 &  4505.0 &  4548.8 &    51.2 &     7.5 & no &0.028 &  Fe4531; & AGE   & -0.397 & AGE   &  0.054 \\
Region05 &  4321.2 &  4340.0 &  4351.2 &    30.0 &    18.8 & no &0.022 &  G4300;H$\gamma_{A}$;H$\gamma_{F}$; & AGE   & -0.392 & AGE   &  0.120 \\
Region04 &  4461.2 &  4470.0 &  4486.2 &    25.0 &     8.8 & no &0.020 &  Ca4455; & AGE   & -0.398 & METAL & -0.277 \\
Region06 &  5162.5 &  5167.5 &  5188.8 &    26.2 &     5.0 & no &0.015 &  Mg$_{2}$;Mg$_{b}$; & AGE   & -0.388 & METAL &  0.166 \\
Region08 &  4636.2 &  4648.8 &  4652.5 &    16.2 &    12.5 & no &0.010 &  C$_{2}$4668; & AGE   & -0.394 & METAL & -0.274 \\
Region11 &  5885.0 &  5891.2 &  5901.2 &    16.2 &     6.2 & no &0.008 &  NaD; & AGE   & -0.374 & AGE   &  0.073 \\
Region13 &  4061.2 &  4071.2 &  4078.8 &    17.5 &    10.0 & no &0.008 &  D$_{n}$(4000)R; & AGE   & -0.398 & METAL & -0.333 \\
Region10 &  4191.2 &  4198.8 &  4202.5 &    11.2 &     7.5 & no &0.007 & \nodata & AGE   & -0.397 & METAL &  0.263 \\
Region15 &  4292.5 &  4300.0 &  4308.8 &    16.2 &     7.5 & no &0.007 &  G4300; & AGE   & -0.398 & METAL &  0.155 \\
Region16 &  4257.5 &  4266.2 &  4273.8 &    16.2 &     8.8 & no &0.007 & \nodata & AGE   & -0.397 & METAL & -0.140 \\
Region09 &  4381.2 &  4387.5 &  4391.2 &    10.0 &     6.2 & no &0.007 &  Fe4383; & AGE   & -0.398 & METAL & -0.174 \\
Region07 &  4913.8 &  4921.2 &  4923.8 &    10.0 &     7.5 & no &0.006 & \nodata & AGE   & -0.393 & METAL & -0.349 \\
Region17 &  5262.5 &  5265.0 &  5272.5 &    10.0 &     2.5 & no &0.005 &  Fe5270; & AGE   & -0.388 & METAL & -0.043 \\
Region20 &  5872.5 &  5876.2 &  5883.8 &    11.2 &     3.8 & no &0.004 &  NaD; & AGE   & -0.375 & METAL &  0.244 \\
Region14 &  4411.2 &  4415.0 &  4420.0 &     8.8 &     3.8 & no &0.004 &  Fe4383; & AGE   & -0.399 & AGE   & -0.097 \\
Region18 &  4311.2 &  4315.0 &  4317.5 &     6.2 &     3.8 & no &0.003 &  G4300; & AGE   & -0.398 & METAL & -0.154 \\
Region19 &  4393.8 &  4395.0 &  4398.8 &     5.0 &     1.2 & no &0.002 &  Fe4383; & AGE   & -0.399 & METAL & -0.139 \\
\hline
\end{tabular}
\\ The \region{s} are sorted by the \pjsumname \ amplitude. If $\lambda_{I} - \lambda_{S} = \lambda_{E} - \lambda_{I}$, the \region \ is called symmetric. AGE stands for stellar age, METAL for stellar metallicity.  The vector \parvector \ is either the AGE or METAL in the \idspace, as that indicated in the columns ``1st mode'' and ``2nd mode''. The vector $u$ is the left singular vector of that \region. The dot product between them, performed in the \idspace, tells how correlated the singular vector is to the corresponding parameter. In both the ``1st mode'' and ``2nd mode'' columns, only the parameter which is most correlated with the singular vector is shown. The extra horizontal line divides the \region{s} with \pjsumname \ less than and larger than \pjsumcutoff.
\label{tab:lickconf}
}
\end{sidewaystable}

\clearpage

\begin{sidewaystable}
{\footnotesize
\caption{Informative wavelength \region{s}, for wavelength range \sdssconfwrange \ at \sdssconfres \ spectral resolution.}
\begin{tabular}{ccccccccccrcr}
\hline
Region & $\lambda_{S}$ & $\lambda_{I}$ & $\lambda_{E}$ & $\lambda_{E} - \lambda_{S}$ & $\lambda_{I} - \lambda_{S}$ & \rotatebox{90}{Symmetric?}
& $\sum p_{\lambda}$ & \lick \ Overlaps & \rotatebox{90}{1st mode}  & \rotatebox{90}{$u^{1} \cdot$ \parvector}  &
\rotatebox{90}{2nd mode}  & \rotatebox{90}{$u^{2} \cdot$ \parvector}  \\
\hline
Region01 &  3450.8 &  3934.5 &  4125.6 &   674.8 &   483.8 & no &0.403 &  H$\delta_{A}$;H$\delta_{F}$;D$_{n}$(4000)L;D$_{n}$(4000)R; & AGE   & -0.390 & METAL & -0.079 \\
Region08 &  4378.0 &  4472.7 &  4574.7 &   196.8 &    94.7 & no &0.063 &  Fe4383;Ca4455;Fe4531; & AGE   & -0.397 & METAL & -0.137 \\
Region03 &  4773.7 &  4861.3 &  4899.5 &   125.8 &    87.6 & no &0.056 &  H$\beta$; & AGE   & -0.393 & METAL & -0.171 \\
Region04 &  4238.2 &  4340.9 &  4377.0 &   138.8 &   102.7 & no &0.049 &  G4300;Fe4383;H$\gamma_{A}$;H$\gamma_{F}$; & AGE   & -0.396 & METAL & -0.177 \\
\hline
Region14 &  4135.1 &  4144.6 &  4237.2 &   102.1 &     9.5 & no &0.028 &  CN$_{1}$;CN$_{2}$;Ca4227; & AGE   & -0.398 & METAL & -0.220 \\
Region12 &  4594.8 &  4651.2 &  4656.5 &    61.7 &    56.4 & no &0.015 &  C$_{2}$4668; & AGE   & -0.395 & METAL & -0.255 \\
Region02 &  6549.4 &  6564.5 &  6576.6 &    27.2 &    15.1 & no &0.012 & \nodata & AGE   & -0.371 & METAL &  0.253 \\
Region09 &  8219.7 &  8229.2 &  8259.5 &    39.8 &     9.5 & no &0.009 & \nodata & AGE   & -0.370 & METAL &  0.290 \\
Region10 &  7166.1 &  7187.6 &  7197.5 &    31.4 &    21.5 & no &0.008 & \nodata & AGE   & -0.375 & METAL &  0.305 \\
Region06 &  5165.9 &  5168.2 &  5179.0 &    13.1 &     2.4 & no &0.006 &  Mg$_{2}$;Mg$_{b}$; & AGE   & -0.388 & METAL &  0.157 \\
Region11 &  5887.3 &  5892.8 &  5902.3 &    14.9 &     5.4 & no &0.006 &  NaD; & AGE   & -0.374 & METAL &  0.093 \\
Region19 &  4914.2 &  4924.4 &  4936.9 &    22.7 &    10.2 & no &0.006 & \nodata & AGE   & -0.393 & METAL & -0.388 \\
Region18 &  8159.4 &  8178.2 &  8182.0 &    22.6 &    18.8 & no &0.005 & \nodata & AGE   & -0.373 & METAL &  0.301 \\
Region17 &  4681.3 &  4686.6 &  4694.2 &    12.9 &     5.4 & no &0.004 &  C$_{2}$4668; & AGE   & -0.393 & AGE   &  0.035 \\
Region07 &  4712.6 &  4714.8 &  4720.2 &     7.6 &     2.2 & no &0.003 &  C$_{2}$4668; & AGE   & -0.393 & METAL &  0.428 \\
Region20 &  5180.2 &  5183.7 &  5189.7 &     9.5 &     3.6 & no &0.003 &  Mg$_{2}$;Mg$_{b}$; & AGE   & -0.387 & METAL &  0.164 \\
Region13 &  5224.5 &  5226.9 &  5231.7 &     7.2 &     2.4 & no &0.003 & \nodata & AGE   & -0.387 & METAL &  0.121 \\
Region05 &  8090.2 &  8092.1 &  8097.7 &     7.5 &     1.9 & no &0.003 & \nodata & AGE   & -0.364 & METAL & -0.121 \\
Region16 &  7274.1 &  7277.4 &  7279.1 &     5.0 &     3.3 & no &0.002 & \nodata & AGE   & -0.375 & METAL &  0.335 \\
Region15 &  5053.0 &  5054.1 &  5056.5 &     3.5 &     1.2 & no &0.002 &  Fe5015; & AGE   & -0.391 & METAL & -0.143 \\
\hline
\end{tabular}
\\ See footnote in \tabname\ref{tab:lickconf}.
\label{tab:sdssconf}
}
\end{sidewaystable}

\clearpage

\begin{sidewaystable}
{\footnotesize
\caption{Informative wavelength \region{s}, for wavelength range \sdssconfwrange \ at \sdssconfres \ spectral resolution, continued.}
\begin{tabular}{cccccccc}
\hline
Region & $\sum p_{\lambda}$ & \lick \ Overlaps & Optical Line Overlaps \\
\hline
Region01 & 0.403 &  H$\delta_{A}$;H$\delta_{F}$;D$_{n}$(4000)L;D$_{n}$(4000)R; &  H$\theta$$\lambda$3799;H$\eta$$\lambda$3836;H$\zeta$$\lambda$3890;CaK$\lambda$3935;CaH$\lambda$3970;H$\epsilon$$\lambda$3971;H$\delta$$\lambda$4103; \\
Region08 & 0.063 &  Fe4383;Ca4455;Fe4531; & \nodata \\
Region03 & 0.056 &  H$\beta$; &  H$\beta$$\lambda$4863; \\
Region04 & 0.049 &  G4300;Fe4383;H$\gamma_{A}$;H$\gamma_{F}$; &  H$\gamma$$\lambda$4342; \\
\hline
Region14 & 0.028 &  CN$_{1}$;CN$_{2}$;Ca4227; & \nodata \\
Region12 & 0.015 &  C$_{2}$4668; & \nodata \\
Region02 & 0.012 & \nodata &  H$\alpha$$\lambda$6565; \\
Region09 & 0.009 & \nodata & \nodata \\
Region10 & 0.008 & \nodata & \nodata \\
Region06 & 0.006 &  Mg$_{2}$;Mg$_{b}$; &  MgI$\lambda$5169;MgI$\lambda$5174; \\
Region11 & 0.006 &  NaD; &  NaI$\lambda$5892;NaI$\lambda$5898; \\
Region19 & 0.006 & \nodata & \nodata \\
Region18 & 0.005 & \nodata & \nodata \\
Region17 & 0.004 &  C$_{2}$4668; & \nodata \\
Region07 & 0.003 &  C$_{2}$4668; & \nodata \\
Region20 & 0.003 &  Mg$_{2}$;Mg$_{b}$; & \nodata \\
Region13 & 0.003 & \nodata & \nodata \\
Region05 & 0.003 & \nodata & \nodata \\
Region16 & 0.002 & \nodata & \nodata \\
Region15 & 0.002 &  Fe5015; & \nodata \\
\hline
\end{tabular}
\\ This table lists the \lick \ indices, and the prominent optical absorption lines, which overlap with the identified \region{s}.
\label{tab:sdssconfcont}
}
\end{sidewaystable}

\clearpage

\epsscale{1.2}

\epsscale{1.0}

\epsscale{1.2}

\epsscale{1.2}

\epsscale{1.0}

\epsscale{1.1}

\clearpage

\epsscale{1.1}

\end{document}